\documentclass[floatfix, preprint, onecolumn, showpacs, showkeys, preprintnumbers,nofootinbib, superscript address,scriptsize]{revtex4-1}

\usepackage{comment}
\usepackage{amssymb}
\usepackage{color}
\usepackage{rotating}
\usepackage{amsmath}
\usepackage{ulem}
\usepackage{overpic}
\usepackage{graphicx}
\usepackage{dcolumn}
\usepackage{graphicx}
\usepackage{bm}
\usepackage{chngpage}
\usepackage{multirow}
\usepackage{slashed}
\usepackage{indentfirst}
\usepackage{booktabs}
\usepackage{amssymb,amsfonts}
\usepackage{amsmath}
\usepackage{color}
\usepackage{hyperref}

\begin{document}
\title{$DK$, $DDK$, and $DDDK$ molecules--understanding the nature of
the $D_{s0}^*(2317)$}
\author{Tian-Wei Wu}
\affiliation{School of Physics and Nuclear Energy Engineering,
  Beihang University, Beijing 100191, China}

\author{Ming-Zhu Liu}
\affiliation{School of Physics and Nuclear Energy Engineering,
  Beihang University, Beijing 100191, China}

    \author{Li-Sheng Geng}
    \email{lisheng.geng@buaa.edu.cn}
    \affiliation{School of Physics and Nuclear Energy Engineering, Beihang University, Beijing 100191, China}
    \affiliation{Beijing Key Laboratory of Advanced Nuclear Materials and
      Physics, Beihang University, Beijing 100191, China}
      \affiliation{School of Physics and Engineering, Zhengzhou University, Zhengzhou, Henan 450001, China}
      
    \author{Emiko Hiyama}
    \email{hiyama@riken.jp}
    \affiliation{Department of Physics, Kyushu University, Fukuoka 819-0395,
      Japan, RIKEN Nishina Center, RIKEN, Wako 351-0198, Japan}
      \affiliation{Nishina Center for Accelerator-Based Science, RIKEN, Wako, 351-0198, Japan}

    \author{Manuel Pavon Valderrama}
    \email{mpavon@buaa.edu.cn}
    \affiliation{School of Physics and Nuclear Energy Engineering,
      Beihang University, Beijing 100191, China}

\begin{abstract}
The $DK$ interaction is strong enough to form a bound state,
the $D_{s0}^*(2317)$.
This in turn begs the question of whether there are bound states
composed of several charmed mesons and a kaon.
Previous calculations indicate that the three-body $DDK$ system
is probably bound, where the quantum numbers are $J^P = 0^{-}$,
$I=\tfrac{1}{2}$, $S = 1$ and $C = 2$.
The minimum quark content of this state is $cc\bar{q}\bar{s}$ with $q=u,d$,
which means that, if discovered, it will be an explicitly exotic tetraquark.
In the present work. we apply the Gaussian Expansion Method to study the $DDDK$ system and show that it binds as well. 
The existence of these three and four body states is rather robust
with respect to the $DD$ interaction and subleading (chiral)
corrections to the $DK$ interaction.
If these states exist, it is quite likely that their heavy quark symmetry counterparts exist as well. 
These three-body $DDK$ and four-body $DDDK$ molecular states could be viewed as counterparts of  atomic nuclei, which are clusters of nucleons bound by the residual strong force, or chemical molecules, which are clusters of atoms bound by the residual electromagnetic interaction. 

\end{abstract}

\maketitle

\section{Introduction}
In 2003 the BaBar collaboration discovered
the $D_{s0}^*(2317)$~\cite{Aubert:2003fg}~\footnote{From now on, we will
simply refer to it as $D_{s0}^*$ unless specified otherwise.}, a strange-charmed scalar meson,
the observation of which was subsequently confirmed
by CLEO~\cite{Besson:2003cp} and Belle~\cite{Krokovny:2003zq}.
Its mass is about $160\,{\rm MeV}$ below the one predicted
for the lightest $c\bar{s}$ scalar state in the naive quark model,
which makes it difficult to interpret the $D_{s0}^*$  
as a conventional $q\bar{q}$ state~\cite{Bardeen:2003kt,Nowak:2003ra,vanBeveren:2003kd,Dai:2003yg,Narison:2003td,Szczepaniak:2003vy,Browder:2003fk,Barnes:2003dj,Cheng:2003kg,Chen:2004dy,Dmitrasinovic:2005gc,Zhang:2018mnm,Terasaki:2003qa,Maiani:2004vq}.

On the other hand, the $D_{s0}^*$ can be easily explained as a
dynamically generated state arising from the Weinberg-Tomozawa (WT)
$DK$ interaction~\cite{Kolomeitsev:2003ac,Hofmann:2003je,Guo:2008gp,Guo:2006fu,Guo:2009ct,Cleven:2010aw,MartinezTorres:2011pr,Torres:2014vna,Yao:2015qia,Guo:2015dha,Albaladejo:2016lbb,Du:2017ttu,Guo:2018kno,Albaladejo:2018mhb,Altenbuchinger:2013gaa,Altenbuchinger:2013vwa,Geng:2010vw,Wang:2012bu,Liu:2009uz,Guo:2018ocg,Guo:2018tjx}.
This has led to the prevailing idea that the $D_{s0}^*(2317)$
is a molecular state, a hypothesis which has been further supported by
a series of Lattice QCD
simulations~\cite{Liu:2012zya,Mohler:2013rwa,Lang:2014yfa,Bali:2017pdv}. For a recent brief summary of all the experimental, lattice QCD, and theoretical supports for such
an assignment, see, e.g., Ref.~\cite{Guo:2019dpg}.

If the $DK$ interaction is strongly attractive, a natural question to ask is
what happens when one adds one extra $D$ meson to
the system~\footnote{It has been shown that the $D\bar{D}^*K$ system
  binds as well in two recent works~\cite{Ma:2017ery,Ren:2018pcd}, though
  the dynamics in these two frameworks are quite different.}.
The answer seems to be that it binds~\cite{SanchezSanchez:2017xtl,MartinezTorres:2018zbl}.
In Ref.~\cite{SanchezSanchez:2017xtl} it was noticed that the $D D_{s0}^*$
system can exchange a kaon near the mass shell, leading to a relatively
long-range attractive Yukawa potential that is strong enough to bind.
This conclusion is left unchanged if one explicitly considers the composite
nature of the $D_{s0}^*$, which simply leads
to more binding~\cite{SanchezSanchez:2017xtl}.
A later, more complete calculation in Ref.~\cite{MartinezTorres:2018zbl}
leads to a binding energy of about 90 ${\rm MeV}$
for the $DDK$ three-body system.
In the present manuscript we revisit the calculation of the $DDK$ bound state
and extend it to the $DDDK$ system by using the Gaussian Expansion Method (GEM),
which  offers a number of advantages compared to previous
studies~\cite{SanchezSanchez:2017xtl,MartinezTorres:2018zbl}.
First, it allows one to calculate directly the density distribution of
the three (four) body system, which then gives a transparent picture
for their spacial distributions.
Second, it has enough flexibility so that one can study the impact of
the existence of a repulsive core.
Indeed, the chiral potential kernel up to the next to leading order
with the low-energy constants determined by the corresponding
lattice QCD data shows that this may indeed
be the case~\cite{Altenbuchinger:2013vwa}.

The outcome of the exploration presented in this work is that
both the $DDK$ and $DDDK$ systems bind, with binding energies of
the order of $65-70$ and $90-100\,{\rm MeV}$ in each case.
While the $DDK$ bound state, owing to its $cc \bar q \bar s $ quark content,
might be produced in experiments in the future, the $DDDK$ bound state
is more likely to be observed on the lattice instead.

This article is organized as follows.
In Sec.~\ref{sec:2B-pot}, we explain how we parametrize and determine
the two-body $DK$ and $DD$ interactions.
In Sec.~\ref{sec:3B-4B}, we explain how to construct the three- and four-body
$DDK$ and $DDDK$ wave functions and solve the corresponding
Schr$\ddot{\rm{o}}$dinger equation using the GEM.
In Sec.~\ref{sec:pred}, we present our predictions
for the $DDK$ and $DDDK$ bound states and discuss their sensitivity
to a series of possible corrections.
Finally, we summarize the results of this manuscript
in Sect.~\ref{sec:conclusions}.

\section{The S-wave $DK$ and $DD$ potentials}
\label{sec:2B-pot}

\begin{table}[htpb]
  \caption{Mass and spin-parity of the $D$, $K$ and $D_{s0}^*(2317)$ mesons.}
  \label{properties}
  \centering
  \begin{tabular}{c c c}
     \hline
     \hline
     Particles& mass(MeV) & $I(J^P)$ \\ 
     \hline
     $D^{\pm}$ & 1869.65 & $\frac{1}{2}(0^-)$ \\ 
     $D^0$ & 1864.83 & $\frac{1}{2}(0^-)$ \\ 
     $K^{\pm}$ & 493.677 & $\frac{1}{2}(0^-)$ \\ 
     $K^0$ & 497.611 & $\frac{1}{2}(0^-)$ \\ 
     $D_{s0}^*(2317)$ & 2317.7 & $0(0^+)$ \\ 
     \hline
     \hline
   \end{tabular}
\end{table}

The calculation of the $DDK$ and $DDDK$ bound states depends on the $DK$ and $DD$
two-body interactions.
While the $DK$ interaction can be well constrained directly from the assumption
that the $D_{s0}^*(2317)$ is a $DK$ bound state, and indirectly
from chiral perturbation theory, the $DD$ interaction
is far from being well determined and we will have
to resort to phenomenological models instead.
In this section we will explain the type of potentials we will use to model
these two-body interactions.

\subsection{The $DK$ interaction}

The most important contribution to the $DK$ interaction is
the WT term between a $D$ meson and a kaon~\footnote{Coupled channel
interactions are small, see, e.g. Ref.~\cite{MartinezTorres:2018zbl}. Therefore, we would work in the single-channel scenario.}.
In the non-relativistic limit we can write this interaction
as a standard quantum mechanical potential,
\begin{eqnarray}
V_{DK}(\vec{q}) = -\frac{C_W(I)}{2 f_\pi^2}
\end{eqnarray}
where the pion decay constant $f_\pi\approx 130$ MeV and $C_W(I)$ represents the strength of the WT interaction,
which is
\begin{eqnarray}
C_W(0) = 2 \quad \mbox{and} \quad C_W(1) = 0 \, ,
\end{eqnarray}
depending on whether we are considering the isospin $I=0$ or $I=1$
configuration of the $DK$ system.
The Fourier-transform of the previous potential in coordinate space is
\begin{eqnarray}
V_{DK}(\vec{r}) = -\frac{C_W(I)}{2 f_{\pi}^2}\,\delta^{(3)}(\vec{r}) \, ,
\end{eqnarray}
which has to be regularized before being used within the Schr\"odinger equation.
A possible choice is to use a local Gaussian regulator of the type
\begin{eqnarray}
  V_{DK}({r}; R_c) = -\frac{C_W(I)}{2 f_{\pi}^2}\,
  \frac{e^{-(r/R_c)^2}}{\pi^{3/2} R_c^3}
\, ,
\end{eqnarray}
where $R_c$ is the cutoff we use to smear the delta function.
For sensible choices of the cutoff, this potential reproduces
the $D_{s0}^*$ pole.
Nowadays we consider the WT interaction
as the leading order (LO) term in the chiral expansion of
the $DK$ potential~\cite{Geng:2010vw,Altenbuchinger:2013vwa}.
In this regard it is interesting to notice that even though
LO chiral perturbation theory (ChPT) indeed indicates that
the $I = 0$ $DK$ interaction in S-wave is attractive,
it happens that the next-to-leading order (NLO) correction is weakly repulsive,
see e.g. Ref.~\cite{Altenbuchinger:2013vwa}.
This motivates the inclusion of a short-range repulsive core
in the $DK$ interaction, as we will explain in the next paragraph.

For the present purposes a more practical approach will be to consider
the $DK$ interaction in a contact-range effective field theory,
in which at LO we have the (already regularized) potential
\begin{eqnarray}
V_{DK}({r}; R_c) = C(R_C)\,\frac{e^{-(r/R_c)^2}}{\pi^{3/2} R_c^3}
\, ,
\end{eqnarray}
with $R_c$ the cutoff and where the $C(R_c)$ is now a running coupling constant.
The differences with a unitarized WT term are
(i) that we let the cutoff $R_c$ to float and
(ii) that we consider the strength of the interaction to run with the cutoff.
In this way by varying the cutoff within a sensible range, for which we choose
$R_c = 1-3\,{\rm fm}$ in this work, we can estimate the uncertainty
in the calculations coming from subleading corrections.
We advance that the cutoff variation will be tiny.
Besides the variation of the cutoff,
we will consider a second method to assess the error in our calculations.
Inspired by the fact that ChPT predicts a repulsive core in the $DK$ interaction
at NLO (as previously mentioned), we can explicitly include
this core in the potential
\begin{eqnarray}
  V_{DK}(\vec{r}; R_c) = C_S \,\frac{e^{-(r/R_S)^2}}{\pi^{3/2} R_S^3} +
  C(R_C)\,\frac{e^{-(r/R_c)^2}}{\pi^{3/2} R_c^3}=C'_S e^{-(r/R_S)^2}+C'_L e^{-(r/R_c)^2}
\, ,
\end{eqnarray}
where $C_S$ is a coupling constant that we set as to provide a repulsive core,
i.e. we take $C'_S> |C'_L|$, and $R_S$ is a second cutoff which fulfills
the condition $R_S < R_c$. For concreteness we take $R_S = 0.5\,{\rm fm}$.

\subsection{The DD interaction}

The $DD$ interaction is not known experimentally,
but there are phenomenological models for it.
Here we will consider the one boson exchange (OBE) potential, which provides
a very simple and intuitive description of the hadron-hadron interactions.
The first qualitatively successful description of the two-nucleon potential
used the OBE model~\cite{Machleidt:1987hj,Machleidt:1989tm},
and the same is true for the first speculations
about the existence of heavy hadron
molecules~\cite{Voloshin:1976ap}.
The particular version of the OBE model that we will use is the one
in Ref.~\cite{Liu:2019stu}, developed for the description of
heavy meson-meson and heavy meson-antimeson systems.

In the particular case of the $DD$ two-body system, the OBE potential involves
the exchange of the $\sigma$, $\rho$ and $\omega$ mesons:
\begin{equation}\label{DD}
  V_{DD}(r; \Lambda)=
  V_{\rho}(r; \Lambda) + V_{\omega}(r; \Lambda) + V_{\sigma}(r; \Lambda)
\end{equation}
where the contribution of each light meson is regularized by means of
a form factor and $\Lambda$ is a cutoff.
The particular contribution of each meson can be written as~\cite{Liu:2019stu}
\begin{eqnarray}
  V_{\sigma}(r; \Lambda) &=&
  - g_{\sigma}^2 \, m_{\sigma} \,
  W_C(m_{\sigma} r, \frac{\Lambda}{m_{\sigma}} ) \, , \\ 
  V_{\rho}(r; \Lambda) &=&
  + \vec{\tau}_1 \cdot \vec{\tau}_2 \, g_{\rho}^2 \, m_{\rho} \,
  W_C(m_{\rho} r, \frac{\Lambda}{m_{\rho}} ) \, , \\
  V_{\omega}(r; \Lambda) &=&
  + g_{\omega}^2 \, m_{\omega} \,
  W_C(m_{\omega} r, \frac{\Lambda}{m_{\omega}} ) \, , 
\end{eqnarray}
where
\begin{equation}
  \label{Y}
  W_C(x, \lambda) = \frac{e^{-x}}{4\pi x} -
  \lambda\,\frac{e^{-\lambda x}}{4\pi \lambda x} -
  \frac{(\lambda^2 - 1)}{2\lambda}\,\frac{e^{-\lambda x}}{4\pi}.
\end{equation}
The masses of the bosons we use are $m_{\rho}=0.770$ GeV, $m_{\omega}=0.780$ GeV,
$m_{\sigma}=0.6$ GeV, and the couplings are $g_{\rho} = g_{\omega} = 2.6$,
$g_{\sigma} = 3.4$.
The cutoff is set by reproducing the $X(3872)$ pole,
yielding $\Lambda = 1.01^{+0.19}_{-0.10}\,{\rm GeV}$~\cite{Liu:2019stu}.
Here for the sake of simplicity we will set the cutoff to $\Lambda = 1.0\,{\rm GeV}$,
where we note that the cutoff dependence is weak.

%

\section{Gaussian Expansion Method to solve the 3-body $DDK$
   and 4-body $DDDK$ systems }
\label{sec:3B-4B}

In this section we briefly explain the
Gaussian Expansion Method (GEM)~\cite{Kamimura:1988zz,Hiyama:2003cu}
as applied to the $DDK$ and $DDDK$ systems.
In the past the GEM has been successfully applied in hypernuclear
as well as heavy-hadron systems.
The focus of the manuscript is on the one hand to confirm the previous
theoretical studies about the existence of a $DDK$ bound state and
to explore whether there are also bound $DDDK$ tetramers.
Regarding the $DDK$ system, it was investigated
in Ref.~\cite{SanchezSanchez:2017xtl} first
as a $D D_{s0}^*$ two-body system, a description which is valid
provided that the size of the $DDK$ trimer is larger than its components
(in particular the $D_{s0}^*$ meson), and second as a genuine
three-body system by solving the Faddeev equations.
In each case the bound state is at about $(50-60)\,{\rm MeV}$
and $(60-100)\,{\rm MeV}$ 
below the $DDK$ threshold, respectively.
Later a more complete study appeared in Ref.~\cite{MartinezTorres:2018zbl},
which uses the method developed by the Valencia group~\cite{MartinezTorres:2007sr,Khemchandani:2008rk,MartinezTorres:2008gy,MartinezTorres:2008kh,MartinezTorres:2009xb,MartinezTorres:2009cw,MartinezTorres:2010zv,MartinezTorres:2011vh,Torres:2011jt} to solve the Faddeev equation~\cite{Faddeev:1960su}
for the $DDK$ system, predicting a bound state
at about $90\,{\rm MeV}$ below the $DDK$ threshold.

\subsection{Three-body $DDK$ system}

\begin{figure}[!h]
  \centering
  \includegraphics[width=15cm]{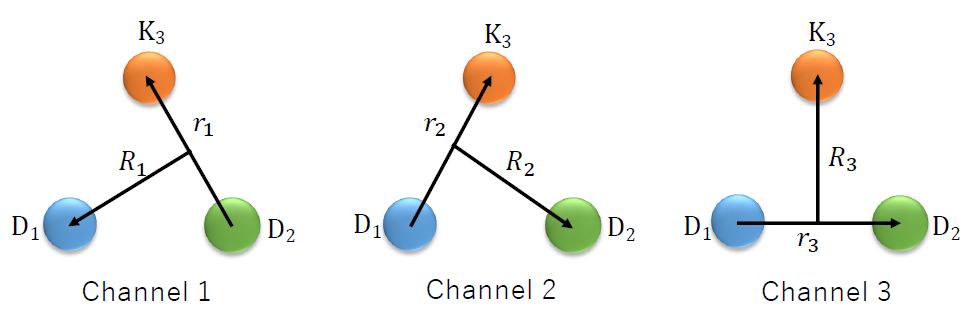}\\
  \caption{The three permutations of the Jacobi coordinates
    for the $DDK$ system}\label{DDK}
\end{figure}

The Schr\"odinger equation of the $DDK$ 3-body system is
 \begin{equation}\label{schd}
   H\Psi_{JM}^{total}=E\Psi_{JM}^{total},
 \end{equation}
with the corresponding Hamiltonian
\begin{equation}\label{hami}
  \hat{H}=\sum_{i=1}^{3}\frac{p_i^2}{2m_i}-T_{c.m.}+\sum_{1=i<j}^{3}V(r_{ij}),
\end{equation}
where $T_{c.m.}$ is the kinetic energy of the center of mass and
$V(r_{ij})$ is the potential between the $i$-th and
the $j$-th particle pair.
The three Jacobi coordinates for the $DDK$ system are shown in Fig.~\ref{DDK}.
The total wave function is a sum of the amplitudes of the three possible
rearrangement of the Jacobi coordinates, i.e. of the channels ($c = 1-3$)
shown in Fig.~\ref{DDK}
  \begin{equation}\label{Sch}
    \Psi_{JM}^{total}=\sum_{c,\alpha}C_{c,\alpha}\,
    \Psi_{JM,\alpha}^{c}(\mathbf{r}_c,\mathbf{R}_c) \, ,
  \end{equation}
  where $\alpha=\{nl,NL,\Lambda,tT\}$ and $C_{c,\alpha}$ are the expansion
  coefficients.
  Here $l$ and $L$ are the orbital angular momenta for the coordinates
  $r$ and $R$,  $t$ is the isospin of the  two-body subsystem in each channel,
  $\Lambda$ and $T$ are the total orbital angular momentum and isospin,
  $n$ and $N$ are the numbers of Gaussian basis function corresponding
  to coordinates $r$ and $R$, respectively.
  For the $DD$ and $DK$ two-body potentials we refer to Sect.~\ref{sec:2B-pot}.
  The eigen energy $E$ and coefficients are determined by
  the Rayleigh-Ritz variational principle.
  Considering that the two $D$ mesons are identical, the total wave function
  should be symmetric with respect to the exchange of the two $D$ mesons,
  which requires that
\begin{equation}\label{exchange}
  P_{12}  \Psi_{JM}^{total}=  \Psi_{JM}^{total},
\end{equation}
and $P_{12}$ is the exchange operator of particles 1 and 2.
The wave function of  each channel has the following form
  \begin{equation}\label{dd}
    \Psi_{JM,\alpha}^{c}(\mathbf{r}_c,\mathbf{R}_c) =
    H_{T,t}^c\otimes[\Phi_{lL,\Lambda}^c]_{JM} \, ,
  \end{equation}
where $H_{T,t}^c$ is the isospin wave function, and $\Phi_{lL,\Lambda}^c$ the spacial wave function. The total isospin wave function reads as
      \begin{equation}\label{isospinwave}
    \begin{split}
          H_{T,t}^{c=1}& =[[\eta_{\frac{1}{2}}(D_2)\eta_{\frac{1}{2}}(K_3)]_{t_1}\eta_{\frac{1}{2}}(D_1)]_{\frac{1}{2}} \, , \\
          H_{T,t}^{c=2}& =[[\eta_{\frac{1}{2}}(D_1)\eta_{\frac{1}{2}}(K_3)]_{t_2}\eta_{\frac{1}{2}}(D_2)]_{\frac{1}{2}} \, , \\
          H_{T,t}^{c=3}& =[[\eta_{\frac{1}{2}}(D_1)\eta_{\frac{1}{2}}(D_2)]_{t_3}\eta_{\frac{1}{2}}(K_3)]_{\frac{1}{2}} \, ,
    \end{split}
  \end{equation}
      where $\eta$ is the isospin wave function of each particle.
      The spacial wave function $\Phi_{lL,\Lambda}^c$ is given in terms of
      the Gaussian basis functions
      \begin{equation}\label{nj}
        \Phi_{lL,\Lambda}^c(\mathbf{r}_c,\mathbf{R}_c)=[\phi_{n_cl_c}^{G}(\mathbf{r}_c)\psi_{N_cL_c}^{G}(\mathbf{R}_c)]_{\Lambda},
      \end{equation}
      \begin{equation}\label{nj}
  \phi_{nlm}^{G}(\mathbf{r}_c)=N_{nl}r_c^le^{-\nu_n r_c^2} Y_{lm}({\hat{r}}_c) \, ,
  \end{equation}
  \begin{equation}\label{nj}
  \psi_{NLM}^{G}(\mathbf{R}_c)=N_{NL}R_c^Le^{-\lambda_n R_c^2} Y_{LM}({\hat{R}}_c) \, .
  \end{equation}
  Here $N_{nl}(N_{NL})$ are the normalization constants of the Gaussian basis
  and the range parameters $\nu_n$ and $\lambda_n$ are given by
  \begin{equation}\label{vn}
  \begin{split}
       \nu_n &=1/r_n^2,\qquad r_n=r_{min}a^{n-1}\quad (n=1,n_{max}) \, , \\
       \lambda_N &=1/R_N^2,\quad R_N=R_{min}A^{N-1}\quad (N=1,N_{max}) \, ,
  \end{split}
  \end{equation}
  in which $\{n_{max},r_{min},a$ or $r_{max}\}$ and  $\{N_{max},R_{min},A$ or
  $R_{max}\}$ are Gaussian basis parameters.
  After the basis expansion, the Schr\"odinger equation of this system
  is transformed into a generalized matrix eigenvalue problem:
\begin{equation}\label{eigenvalue problem}
  [T_{\alpha \alpha'}^{ab}+V_{{\alpha \alpha'}}^{ab}-EN_{\alpha \alpha'}^{ab}]\,
  C_{b,\alpha'} = 0
  \, .
\end{equation}
Here, $T_{\alpha \alpha'}^{ab}$ is the kinetic matrix element, $V_{\alpha \alpha'}^{ab}$ is the potential matrix element and $N_{\alpha \alpha'}^{ab}$ is the normalization matrix element.

The quantum numbers of all the allowed configurations are determined by angular momentum conservation, isospin conservation, parity conservation, and Bose-Einstein statistics.  Given that
we only consider $S$-wave interactions, and only the $DK$ interaction in $I=0$ is dominant, we obtain the allowed configurations shown in Table \ref{ddkconf}. The $DDK$ system that we are interested in has isospin 1/2 and spin parity $0^-$. 

\begin{table}[h]
	\caption{Quantum numbers of different Jacobi coordinate channels $(c=1-3)$ of the $DDK$ $I(J^P)=\frac{1}{2}(0^-)$ state. Note that channel 1 and channel 2 are the same.}\label{ddkconf}
	\centering
	\begin{tabular}{c c c c c c c c}
		\hline\hline
		c & $l$ & $L$ & $\Lambda$ & $t$ & $T$ & $J$ & $P$\\
		\hline
		1(2) & 0 & 0  & 0 & 0 & $\frac{1}{2}$ & 0& $-$ \\
		1(2) & 0 &0 & 0 & 1 & $\frac{1}{2}$ & 0& $-$ \\
		3 & 0 & 0 & 0 & 1 &$\frac{1}{2}$  & 0 & $-$ \\
		\hline\hline
	\end{tabular}
\end{table}
\subsection{Four-body $DDDK$ system}

\begin{figure}[t]
	\centering
	\includegraphics[width=12cm]{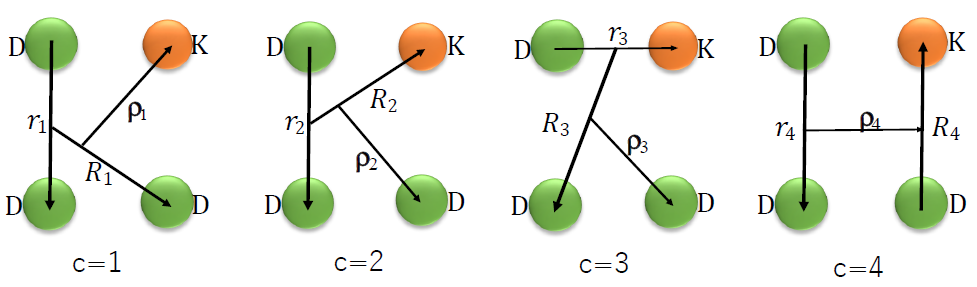}\\
	\caption{Jacobi coordinates for the rearrangement channels ($c=1-4$) of  the $DDDK$ system. The three $D$ mesons are to be symmetrized.}\label{DDDK}
\end{figure}

A generic four-body system has 18 Jacobi coordinates.
In the $DDDK$ system, owing to the fact that there are three identical
$D$ mesons, the possible configurations of the Jacobi coordinates
reduce to three K-type channels and one H-type channel,
see Fig.[\ref{DDDK}].
There are 4 identical Jacobi coordinates for each K-type channel
and 6 identical Jacobi coordinates for the H-type channel.
The total wave function of this $DDDK$ system is 
  \begin{equation}\label{Sch}
\Psi_{I(J^P)}^{total}=\sum_{c,\alpha}A_{c,\alpha}\Psi_{\alpha}^{c}(\bm{r}_c,\bm{R}_c,\bm{\rho}_c), \qquad c=1-18 \, ,
\end{equation}
and the wave function in each Jacobi channel reads
  \begin{equation}\label{dd}
\Psi_{\alpha}^{c}(\bm{r}_c,\bm{R}_c,\bm{\rho}_c)=H_{t,T,I}^c\otimes\Phi_{lL\lambda,\sigma\Lambda}^{c,JP} \, .
\end{equation}
  Here $t,T,I$ are the isospin of the coordinates $r, R$ and $\rho$
  in each channel; $l, L$ and $\lambda$ are the orbital angular momenta
  for the coordinates $r,R$ and $\rho$, while $\sigma$ is the coupling
  of $l$ and $L$, $\Lambda$ is the coupling of $\sigma$ and $\lambda$,
  and $J, P$ is the total angular momentum and parity.
  The Gaussian basis and parameters are in the same form as those
  in the  3-body system, which are
    \begin{equation}\label{nj}
\Phi_{lL\lambda,\sigma\Lambda}^c=[\phi_{n_cl_c}^{G}(\bm{r}_c)\psi_{N_cL_c}^{G}(\bm{R}_c)]_{\sigma_c}\varphi_{\nu_c\lambda_c}^{G}(\bm{\rho}_c)]_{\Lambda} \, ,
\end{equation}
\begin{equation}\label{nj}
\phi_{nlm}^{G}(\bm{r}_c)=N_{nl}r_c^le^{-\nu_n r_c^2} Y_{lm}({\hat{r}}_c) \, ,
\end{equation}
\begin{equation}\label{nj}
\psi_{NLM}^{G}(\bm{R}_c)=N_{NL}R_c^Le^{-\lambda_N R_c^2} Y_{LM}({\hat{R}}_c) \, ,
\end{equation}
\begin{equation}\label{nj}
  \varphi_{\nu\lambda\mu}^{G}(\bm{\rho}_c)=
  N_{\nu\lambda}\rho_c^{\lambda}e^{-\omega_{\nu} \rho_c^2} Y_{\lambda\mu}({\hat{\rho}}_c) \, .
\end{equation}
Here $N_{nl}(N_{NL})$ are the normalization constants of
the Gaussian basis and the range parameters
$\nu_n$, $\lambda_n$ and $\omega_{\nu}$
are given by
\begin{equation}\label{vn}
\begin{split}
\nu_n &=1/r_n^2,\qquad r_n=r_{min}a^{n-1}\quad (n=1,n_{max}) \, , \\
\lambda_N &=1/R_N^2,\quad R_N=R_{min}A^{N-1}\quad (N=1,N_{max}) \, , \\
\omega_{\nu} &=1/\rho_{\nu}^2,\qquad \rho_{\nu}=\rho_{min}\alpha^{\nu-1}
\quad (\nu=1,\nu_{max}) \, .\\
\end{split}
\end{equation}
Since we are considering only $S$-wave interactions,
we have $J=l=L_\lambda=\sigma=\Lambda=0$, and the parity is $+$.
The procedure to  determine the allowed configurations for the $DDDK$ system
is the same as the $DDK$ case.
The  4-body $DDDK$ configurations are shown in Table.\ref{DDDK:Conf}.
\begin{table}[t]
  \caption{Quantum numbers of different Jacobi coordinate
    channels $(c=1-4)$ of the $DDDK$ $I(J^P)=1(0^+)$ state.
    The identical channels have the same configuration.
    The number in the brackets denotes the alternative value. }\label{DDDK:Conf}
	\centering
	\begin{tabular}{c c c c c c c c c c c}
		\hline\hline
		c & $l$ & $L$ & $\lambda$ & $\sigma$ & $L$ & $t$ & $T$ & $I$ & $J$ & $P$ \\
		\hline
		1 & 0 & 0 & 0 & 0 & 0  & 1 & $\frac{1}{2}(\frac{3}{2})$ & 1 & 0 & $+$ \\
		2 & 0 & 0 & 0 & 0 & 0 & 1 & $\frac{1}{2}(\frac{3}{2})$ & 1 & 0 & $+$ \\
		3 & 0 & 0 & 0 & 0 & 0 & 0(1) & $\frac{1}{2}(\frac{3}{2})$ & 1 & 0 & $+$ \\
		4 & 0 & 0 & 0 & 0 & 0 & 1 & 0(1) & 1 & 0 & + \\
		\hline\hline
	\end{tabular}
\end{table}

\section{Predictions}
\label{sec:pred}

In this section we discuss the predictions
we make for the $DDK$ and $DDDK$ bound states.
With the two-body inputs of Sect.~\ref{sec:2B-pot} and
the three(four)-body configurations detailed in Sect.~\ref{sec:3B-4B},
we can predict the existence of $DDK$ and $DDDK$ bound states.
The outcome is that the $DDK$ trimer will bind by about $70\,{\rm MeV}$
and the $DDDK$ tetramer by about $100\,{\rm MeV}$, with variations
of a few ${\rm MeV}$ at most, stemming from the uncertainties
in the $DK$ and $DD$ potentials.

\subsection{Solving the $DDK$ and $DDDK$ systems}

\begin{figure}[!h]
	\centering
	\includegraphics[width=5cm]{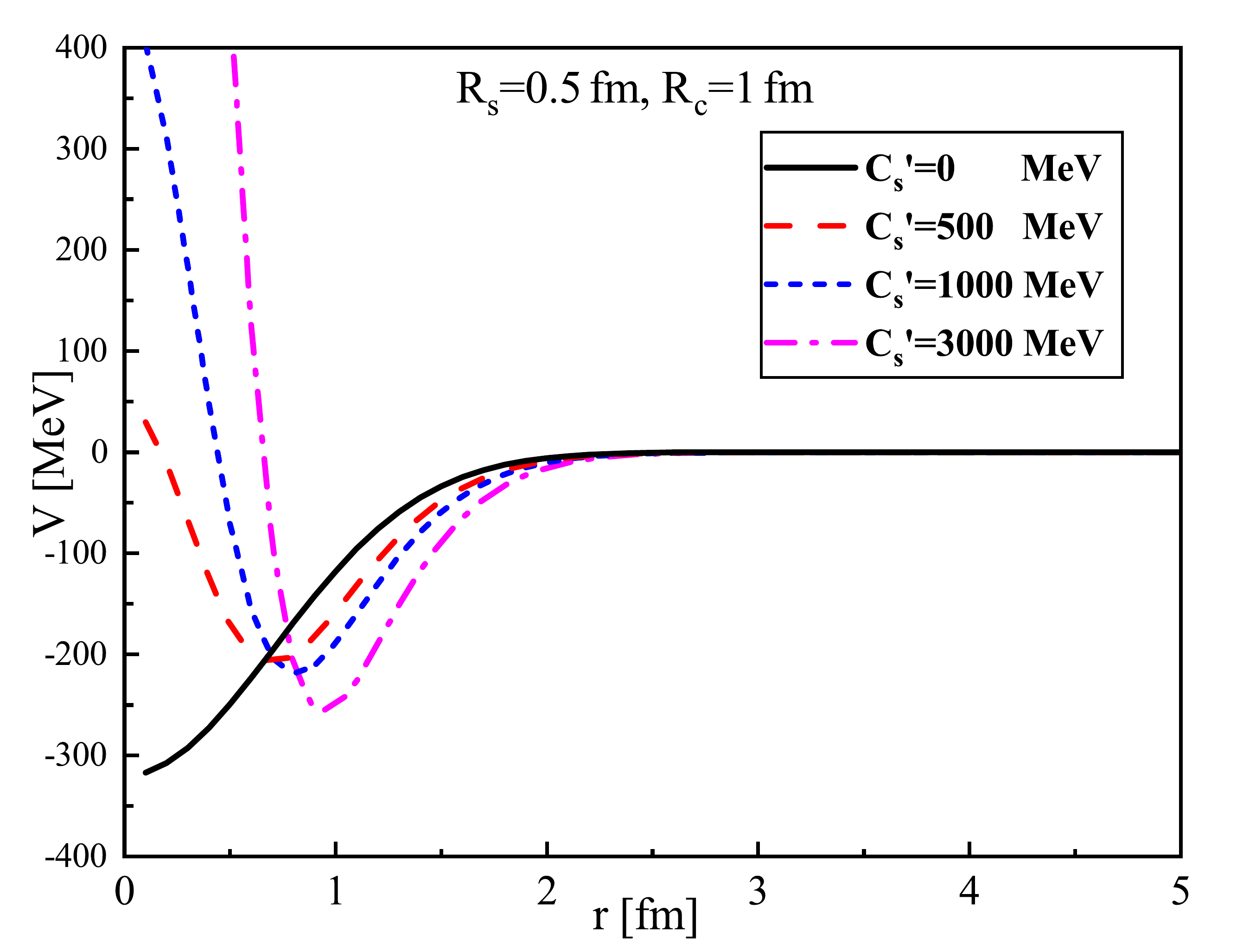} \includegraphics[width=5cm]{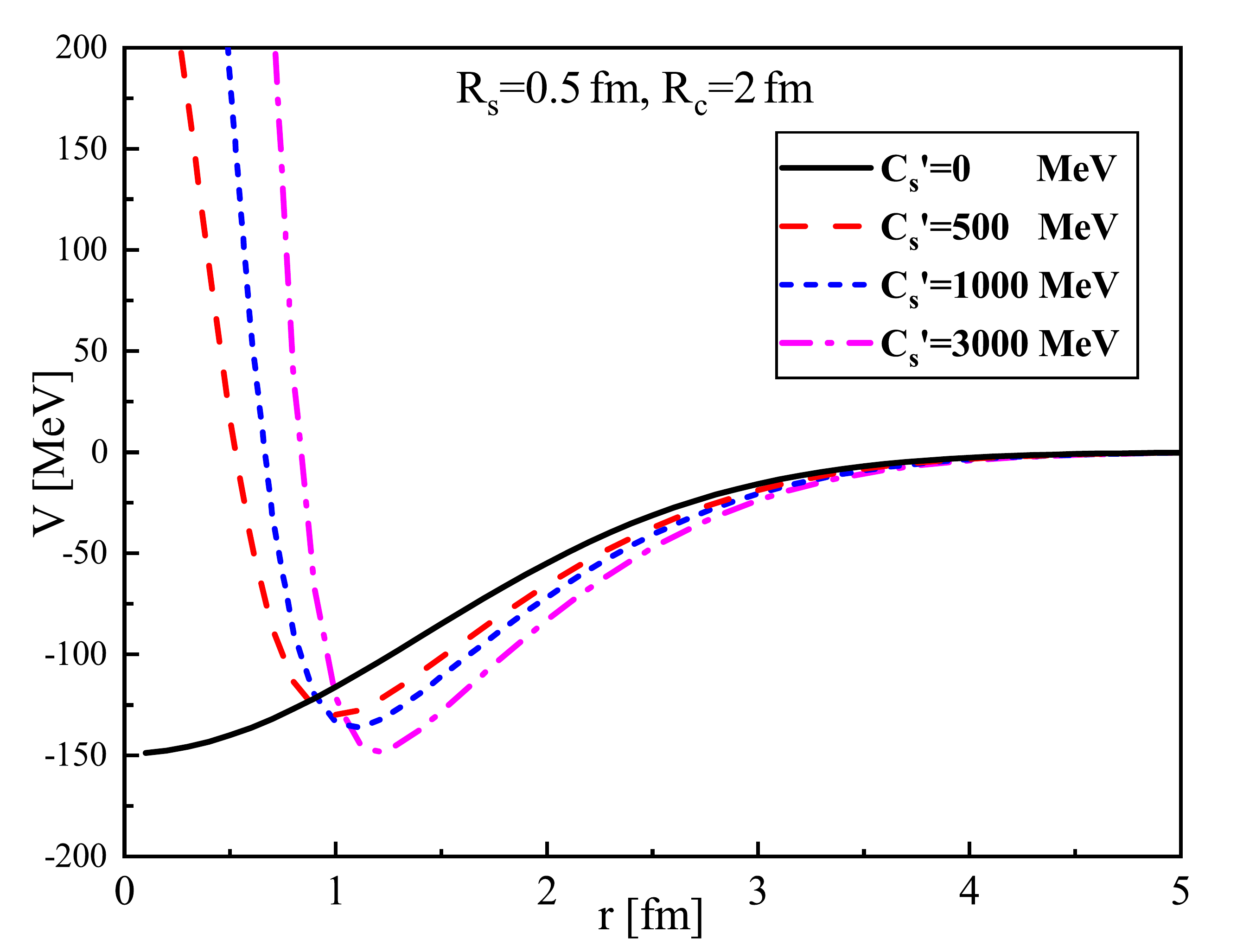} \includegraphics[width=5cm]{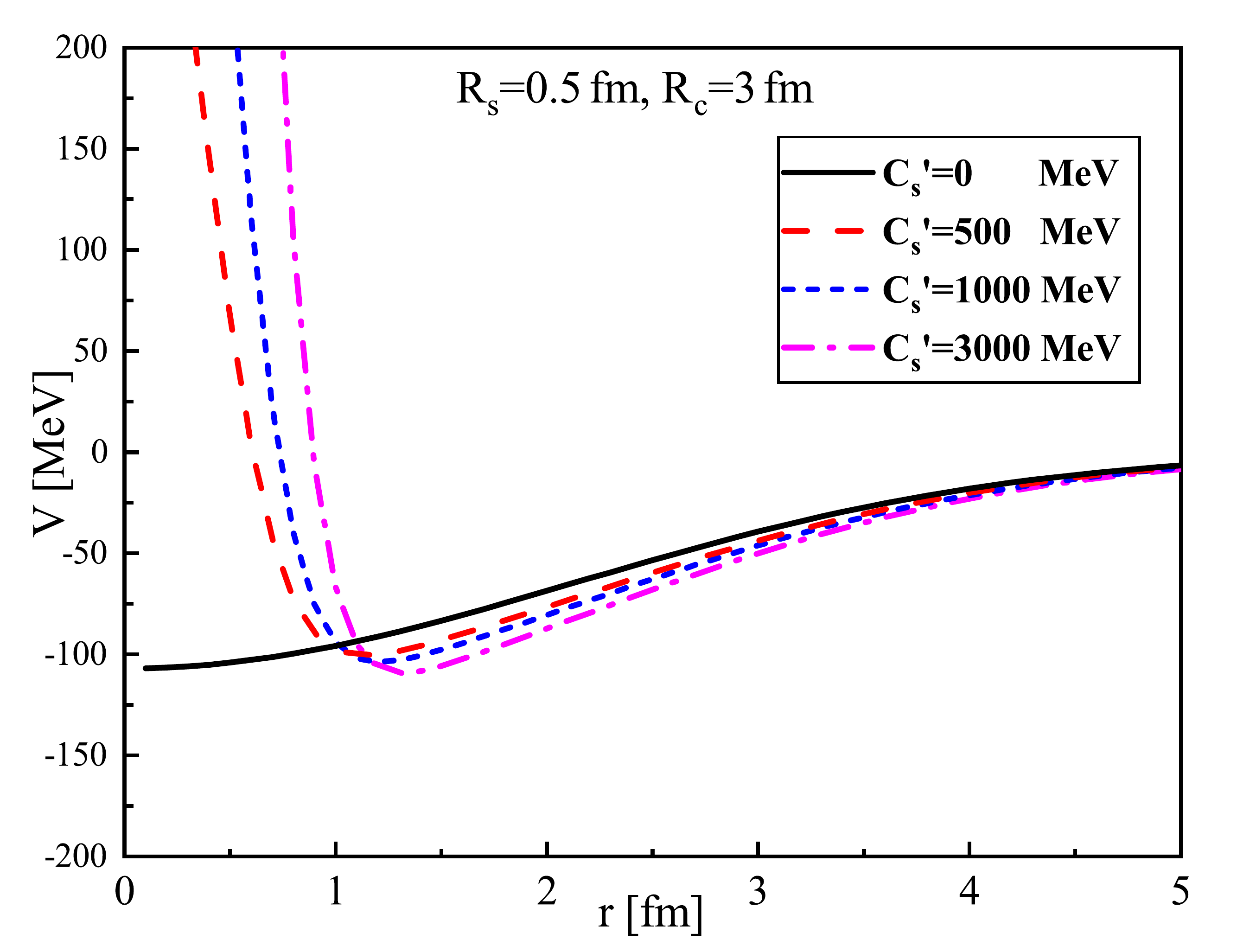}\\
	\caption{ Isospin $t=0$ $DK$ potential as a function of the distance between $D$ and $K$ for different $R_S$, $R_c$, and $C'_S$. The coupling $C'_L$ in each case is determined by reproducing the $D_{s0}^*$. }\label{potential}
\end{figure}

\begin{figure}[h!]
	\centering
	\includegraphics[width=5cm]{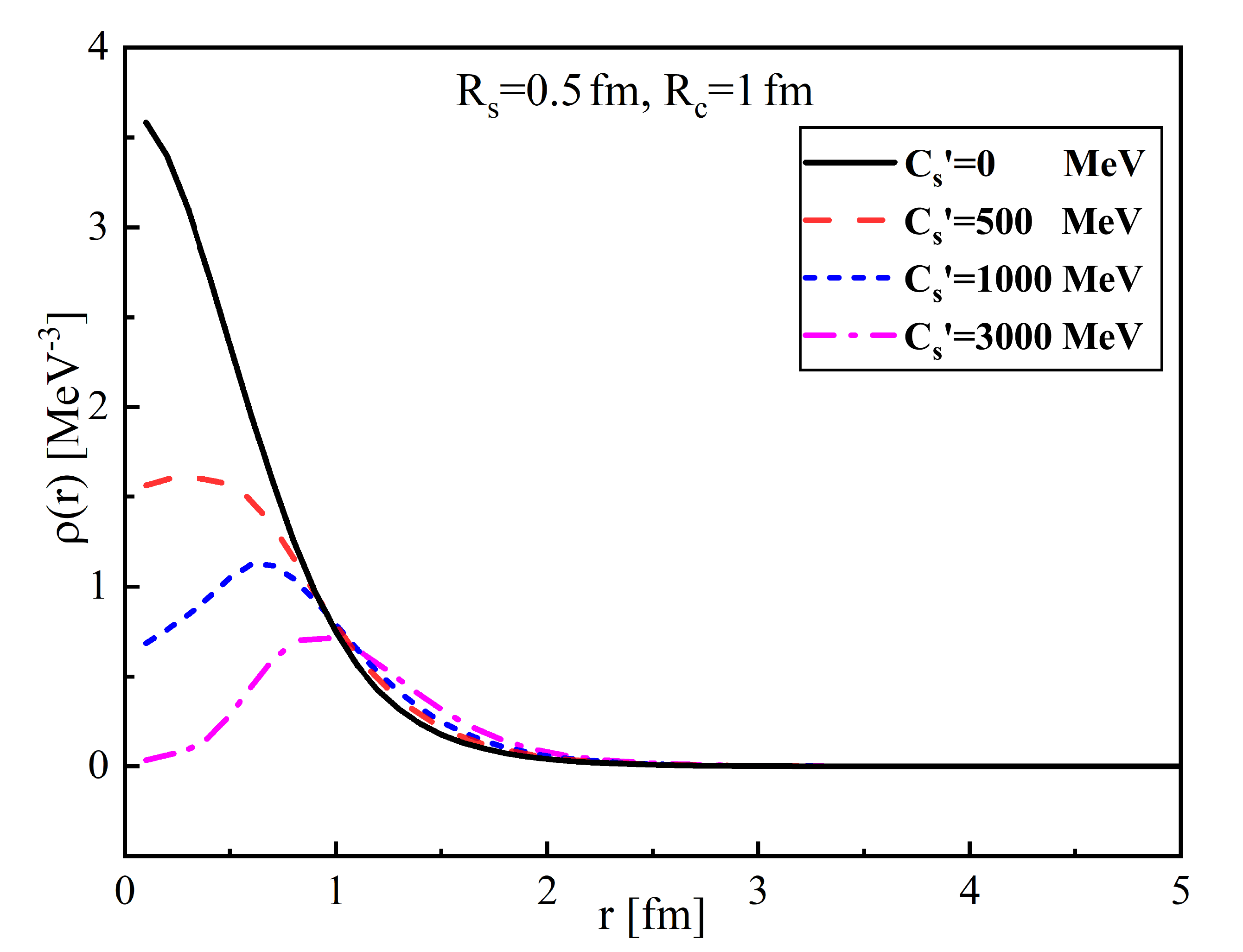} \includegraphics[width=5cm]{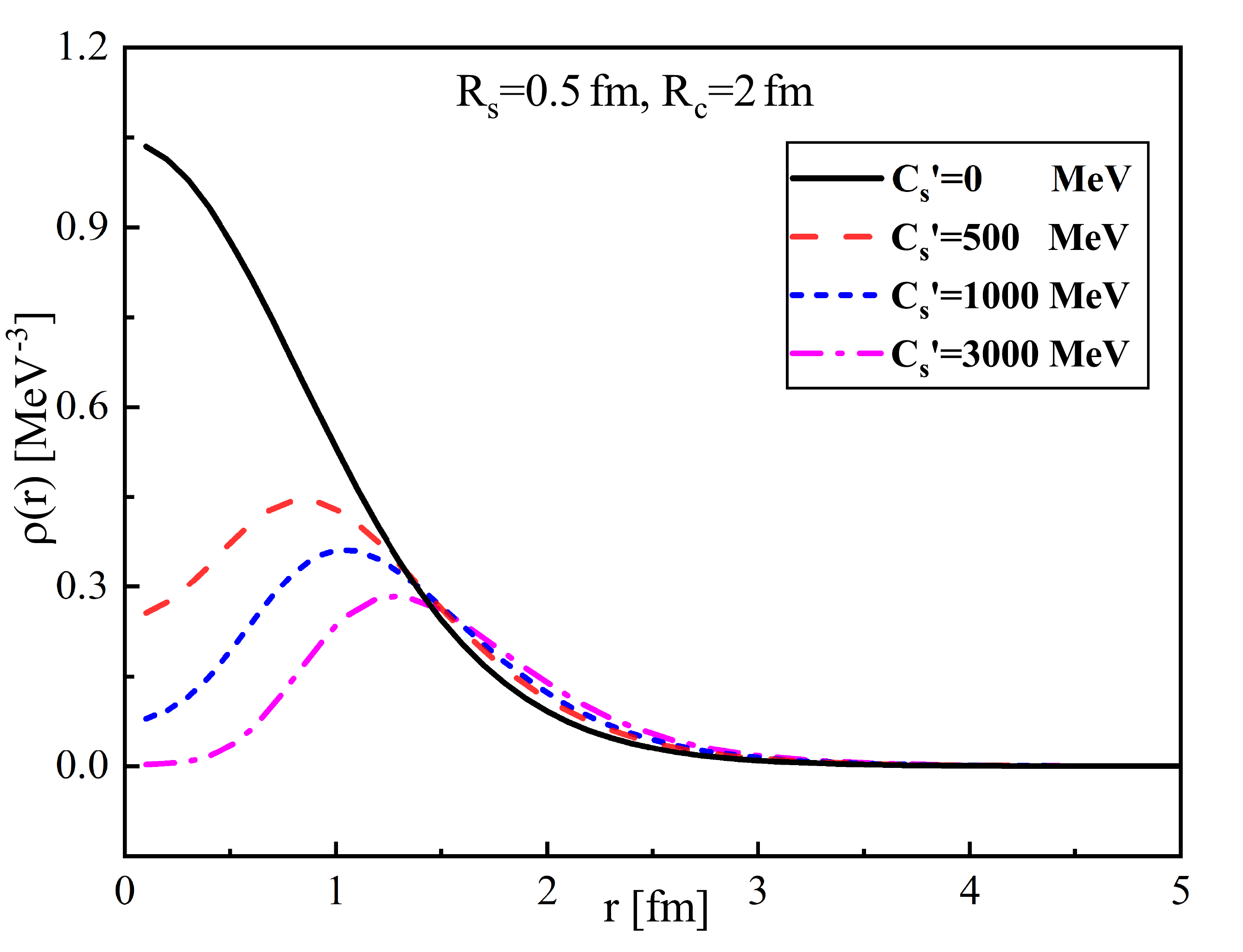} \includegraphics[width=5cm]{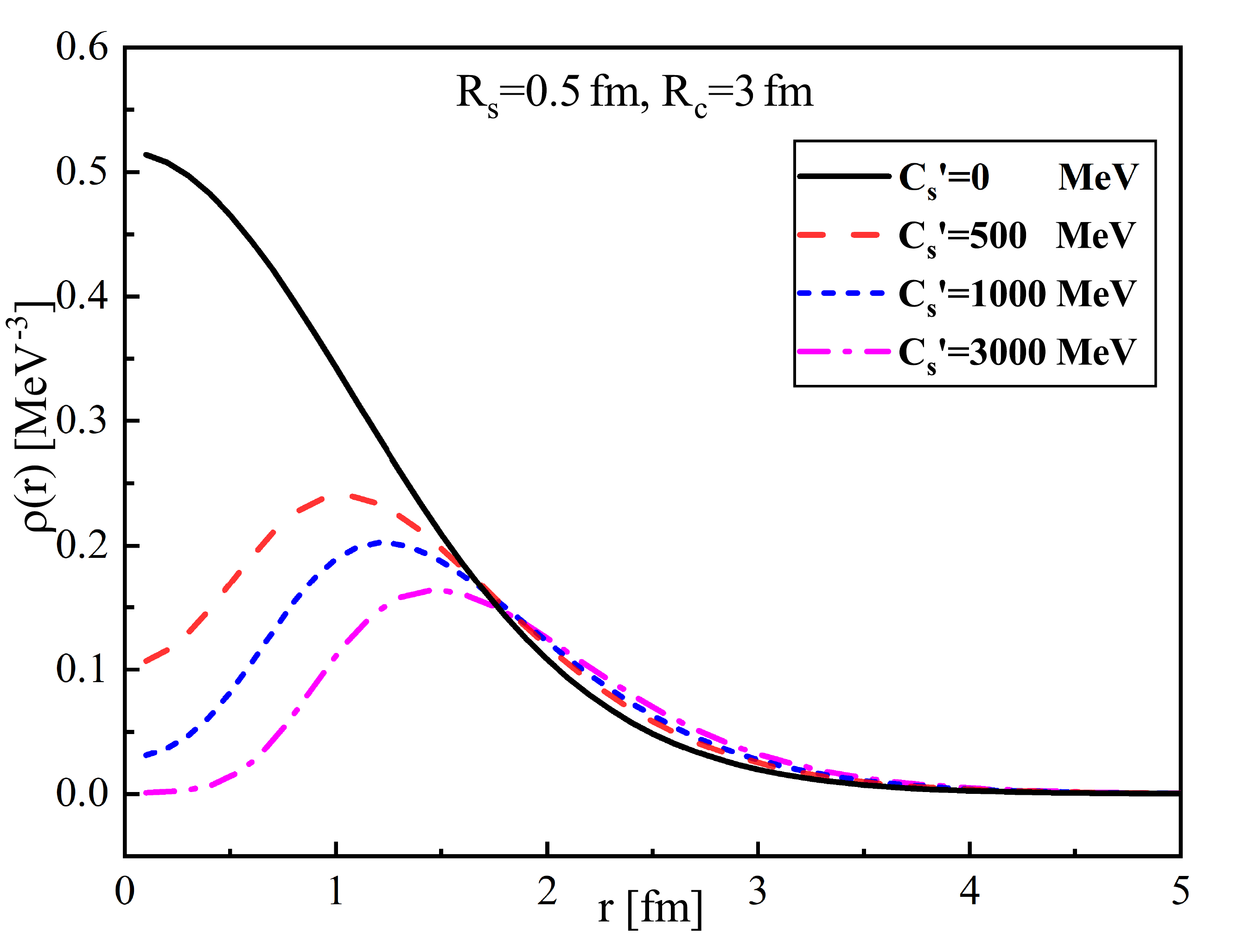}\\
	\caption{Density profile of the $DK$ molecule corresponding to the potentials of Fig.~\ref{potential}.}\label{pdc}
\end{figure}

The two basic input blocks for the calculation of the $DDK$ and $DDDK$ systems
are the $DK$ and $DD$ interactions, of which the $DK$ one is
the most important factor when it comes to binding.
The $DK$ potential contains the running coupling $C(R_c)$ and the cutoff $R_c$,
where $R_c = (1-3)\,{\rm fm}$ and $C(R_c)$ is determined from the condition
of reproducing the well-known $D_{s0}^*(2317)$ as a $DK$ bound state
with a binding energy of $45\,{\rm MeV}$.
In addition there are two additional parameters, the coupling $C_S$
and the short-range radius $R_s = 0.5\,{\rm fm}$,
which are used to estimate the uncertainties
in the $DK$ potential.
We study three combinations of $R_S$ and $R_c$, which can be consulted
in Table \ref{Results:BE}, where we also list the values of
the couplings $C_R$ and $C(R_c)$ and the binding energies
of the $DDK$ and $DDDK$ systems.
The different potentials investigated are shown in Fig.~\ref{potential}
and the probability density distributions of the $DK$ pair corresponding to
the potentials are shown in Fig.~\ref{pdc}.
\begin{table}[!h]
  \caption{Binding energies (in units of MeV) of $DDK$ and $DDDK$
    systems with and without the $DD$ interaction for different combinations of $R_S$, $R_c$, $C'_S$, and $C'_L$. The couplings are in units of MeV.}\label{Results:BE}
	\centering
	\begin{tabular}{c c c c c c c }
		\hline
		\hline
		$C'_S$ &$C'_L$ & $E_2$ &$E_3$(only $V_{DK})$ &$E_3(V_{DK}+V_{DD})$&$E_4($only $ V_{DK})$ &$E_4(V_{DK}+V_{DD})$ \\
		\hline
		&&$R_S=0.5$ fm&& $R_c=1$ fm&&  \\
		\hline
		0&  $-320.1$  & $-45.0$ & $-65.8$ & $-71.2$& $-89.4$ & $-106.8$ \\
		$500$  & $-455.4$ & $-45.0$ & $-65.8$ & $-70.4$& $-89.2$ & $-103.5$ \\
		$1000$ & $-562.6$ & $-45.0$ & $-65.7$ & $-69.7$ & $-88.8$ & $-101.4$\\
		$3000$ & $-838.7$ & $-45.0$ & $-65.0$ & $-68.4$& $-87.0$ & $-97.3$ \\
		
		\hline
		&&$R_S=0.5$ fm && $R_c=2$ fm&& \\
		\hline
		0&  $-149.1$  & $-45.0$ & $-66.0$ & $-68.8,-45.1$& $-88.7,-66.3$ & $-97.6,-70.7$  \\
		$500$ & $-178.4$ & $-45.0$ & $-65.9$ & $-68.2,-45.5$& $-88.5,-66.7$ & $-95.5,-70.9$\\
		$1000$ & $-195.0$ & $-45.0$ & $-65.8,-45.2$ & $-67.9,-45.8$ & $-88.2,-66.9$ & $-94.5,-71.2$\\
		$3000$ & $-225.9$ & $-45.0$ & $-65.3,-45.6$ &  $-67.2,-46.6$ & $-87.0,-67.0$ & $-92.6,-71.7$\\
		
		\hline
		&&$R_S=0.5$ fm && $R_c=3$ fm&& \\
		\hline
		0&$-107.0$ & $-45.0$ & $-66.2,-47.3$ & $-68.0,-48.3$& $-88.8,-70.2$ & $-94.4,-74.3$\\
		$500$ & $-119.4$ & $-45.0$ & $-66.2,-48.2$ & $-67.7,-49.3$& $-88.7,-71.0$ & $-93.2,-74.8$ \\
		$1000$ & $-125.6$ & $-45.0$ & $-66.1,-48.7$ & $-67.5,-49.8$& $-88.4,-71.3$ & $-92.5,-75.2$\\
		$3000$ & $-136.2$ & $-45.0$ & $-65.8,-49.4$ &$-67.1,-50.7$ & $-87.6,-71.7$ & $-91.4,-75.7$ \\
		\hline
		\hline
	\end{tabular}
\end{table}

\begin{table}[!h]
  \caption{Root mean square (RMS) radius (in units of fm) of $DK$
    and $DDK$ systems, the expectation values  (in units of MeV) of
    the kinetic term, $DK$ and $DD$ interactions   with various parameters $R_S$. $R_c$, $C'_S$, and $C'_L$. The couplings are in units of MeV.}\label{Results:RMS}
	\centering
	\begin{tabular}{c c c c c c c c}
		\hline
		\hline
		$C'_S$ &$C'_L$ & $r_2(DK)$ &$r_3(DK)$ &$r_3(DD)$&$<T>$ &$<V_{DK}>$ &$<V_{DD}>$  \\
		\hline
		&&$R_S=0.5$ fm && $R_c=1$ fm&&&  \\
		
		\hline
		0&  $-320.1$  & 1.28 &$1.32$&1.36&$124.37$ & $-189.61$ &$-5.98$  \\
		$500$  & $-455.4$& 1.39 &$1.44$&1.47& $99.51$ & $-164.83$&$-5.03$  \\
		$1000$ & $-562.6$& 1.46 &$1.53$&1.54& $91.43$ & $-156.67$&$-4.51$ \\
		$3000$ & $-838.7$& 1.61 &$1.69$&1.68& $93.24$ & $-157.80$ &$-3.82$\\
		
		\hline
		&&$R_S=0.5$ fm & &$R_c=2$ fm&&&  \\
		\hline
		0&  $-149.1$  & 1.74 &$1.80$&1.80&$60.20$ & $-125.74$ &$-3.23$\\\
		$500$ & $-178.4$& 1.91 & $1.98$&1.96& $51.00$ & $-116.59$& $-2.64$\\
		$1000$ & $-195.0$& 1.99 &$2.07$&2.04& $50.63$ & $-116.12$&$-2.43$\\
		$3000$ & $-225.9$& 2.13 &$2.22$&2.15& $53.61$ & $-118.59$&$-2.24$\\
		
		\hline
		&&$R_S=0.5$ fm && $R_c=3$ fm&&&  \\
		\hline
		0&$-107.0$ & $2.13$ &$2.19$&2.17& $39.49$ & $-105.35$&$-2.13$\\
		$500$ & $-119.4$ & $2.31$ &$2.38$&2.34& $34.80$ & $-100.73$ &$-1.77$\\
		$1000$ & $-125.6$& $2.37$ &$2.47$&2.42& $34.90$ & $-100.77$ &$-1.65$\\
		$3000$ & $-136.2$& $2.53$ &$2.61$&2.53& $36.66$ & $-102.24$ &$-1.54$\\
		\hline\hline
	\end{tabular}
\end{table}

A few comments about the results of Table \ref{Results:BE} are in order.
The first thing we notice is that the impact of the $DD$ interaction is mild.
It makes the $DDK$ and $DDDK$ systems more bound, but only
by a few ${\rm MeV}$.
This is a bit relieving as the $DD$ interaction is not well known.
The second interesting observation is that the existence of
the $DDK$ and $DDDK$ bound states is rather robust
with respect to the likely existence of
a short-range repulsive core.
In other words, the existence of the $DDK$ and $DDDK$ bound states is
almost guaranteed as long as the $D_{s0}^*$ is dominantly
a $DK$ bound state (we will later check that this will still
be the case even if the $D_{s0}^*$ is a compact
${\bar c} s$ state).
The third observation is that as the range of the attraction becomes larger,
two bound state solutions appear instead of one, with the deepest
bound one becoming slightly shallower.

In Table \ref{Results:RMS} we show the root mean square (RMS) radius of
the $DK$ and $DDK$ systems as well as the expectation values
of the kinetic and potential terms.
The RMS radius of the $D_{s0}^*$, which ranges from $1.2$ to
$2.6$ fm, increases with the cutoff $R_c$ and
with the coupling $C_S$ of the short-range
repulsive core.
In the $DDK$ system, the RMS radius of the $DK$ pair is slightly
larger than its counterpart in the $D_{s0}^*$.
The RMS radius of the $DD$ system also increases if we increase the cutoff
$R_c$ or the coupling $C_S$.
We notice that the geometry of the $DDK$ system is more or less of
a proper triangle, which agrees qualitatively
with the findings of Ref.~\cite{MartinezTorres:2018zbl}.
From the last two columns of Table \ref{Results:RMS},
it is clear that the $DD$ interaction is weakly attractive,
accounting for only a few ${\rm MeV}$ of the total potential energy.

%

\subsection{Solving the $DDDK$ system as an equivalent  $DDD_{s0}^*$ system }

If the separation of the $DK$ pair within the $DDK$ trimer and $DDDK$ tetramer
is comparable to or larger than the expected size of the $D_{s0}^*$,
in a first approximation it will be possible to treat the $D_{s0}^*$
as a point-like particle, with its compound structure providing
subleading corrections to this point-like approximation.
From Table \ref{Results:RMS} we can see that the RMS of the $DK$ subsystem
in the $DDK$ and $DDDK$ systems is similar to that of
the $D_{s0}^*$ as a $DK$ molecule.
In this regard we notice that in Ref.~\cite{SanchezSanchez:2017xtl}
the $D_{s0}^*$ is approximated as point-like, where the interaction
between the $D$ and $D_{s0}^*$ is mediated by one kaon exchange
and is strong enough to form a bound state.
This $D D_{s0}^*$ molecule is predicted to be $50-60\,{\rm MeV}$ below
the $DDK$ threshold, to be compared with $65\,{\rm MeV}$ when we
consider it as a genuine $DDK$ three-body state and ignore
the $DD$ interaction (see Table \ref{Results:BE}).
This indicates that the predictions of the point-like approximation are
reasonably good (for such a simple approximation) and that
the compound structure of the $D_{s0}^*$ provides
additional attraction.
In the following lines we will extend the ideas of Ref.~\cite{SanchezSanchez:2017xtl}
to the $DDDK$ tetramer, i.e. we will treat it as a three-body $DDD_{s0}^*$
system where the $D_{s0}^*$ is assumed to be a compact meson.
To do this, we first reproduce the two-body calculation
of Ref.~\cite{SanchezSanchez:2017xtl}, but in coordinate space,
and then study the three-body $DDD_{s0}^*$ system
using the GEM.

The interaction of $DD_{s0}^*$ is attractive and
reads as\begin{equation}\label{oke}
V_{OKE}(\vec{q})=-h^2\frac{\omega_K^2}{f_{\pi}^2}\frac{1}{\mu_K^2+\vec{q}^2},
\end{equation}
where $\omega_K=m_{D_{s0}^*} -m_D$ and the effective kaon mass
$\mu_K=\sqrt{m_K^2-\omega_K^2}$.
As in Ref.~\cite{SanchezSanchez:2017xtl}, we take $h=0.7$ and $f_{\pi}=130$ MeV.
We regularize the potential by multiplying it with a dipole form factor of
the type:
\begin{equation}\label{ff}
F_D(q^2) =\frac{(\Lambda^2-m_K^2)^2}{(\Lambda^2-q^2)^2} \, .
\end{equation}
After the inclusion of this form factor,
the $DD_{s0}^*$ potential in coordinate space reads
\begin{equation}\label{coordDDs0}
V_{DD_{so}^*}(r)=-h^2\frac{\omega_K^2}{f_{\pi}^2}\left(\frac{e^{-\mu_{K}r}}{4\pi r}-\frac{e^{-\Lambda' r}}{4\pi r}-\frac{(\Lambda'^2-\mu_K^2)e^{-\Lambda' r}}{8\pi \Lambda'}\right) \, ,
\end{equation}
where we define $\Lambda'$ as
\begin{equation}\label{Lambdap}
\Lambda'^2=\Lambda^2-q_0^2=\Lambda^2-\omega_K^2 \, .
\end{equation}
Using the above $DD_{s0}^*$ potential and the $DD$ potential provided
by the OBE model, we can check whether the three-body $DDD_{s0}^*$
system binds.
The binding energies we obtain with different cutoffs are tabulated
in Table \ref{tab:DDDs0}.
\begin{table}[!h]
	\caption{Binding energies (in units of MeV) of $DD_{s0}^*$ and ${DDD_{s0}^*}$ systems with different cutoff $\Lambda'$ (in units of GeV).\label{tab:DDDs0}}
	\centering
	\begin{tabular}{c c c c }
		\hline
		\hline
		$\Lambda'$ & $B_{DD_{s0}^*}$ & $B_{DDD_{s0}^*}$(only $V_{DD_{s0}^*}$)& $B_{DDD_{s0}^*}$($V_{DD}+V_{DD_{s0}^*}$)\\
		\hline
		0.8 &$ -5.1 $& $-11.5$ &$-13.9$\\
		1.0 & $-8.5$ & $-18.9$ &$-22.5$\\
		1.2 & $-11.7$ & $-25.8$ &$-30.3$\\
		1.4 & $-14.5$ & $-31.9$ &$-37.2$\\
		1.6 & $-17.0$ & $-37.2$ &$-43.3$\\
		\hline
		\hline
	\end{tabular}
\end{table}

With the effective cutoff $\Lambda'$ ranging from $0.8-1.6$ GeV,
the results of Table.\ref{tab:DDDs0} indicate that the $DDD_{s0}^*$
bound state is located about $(65-90)\,{\rm MeV}$ below
the $DDDK$ threshold.
This is to be compared with $100\,{\rm MeV}$ for the full four-body calculation,
see Table.\ref{Results:BE} for details.
That is, as happened with the $DD_{s0}^*$ / $DDK$ system, the approximation
that the $D_{s0}^*$ is a compact state results in underbinding
for the $DDD_{s0}^*$ / $DDDK$ system, but not much.

\section{Summary}
\label{sec:conclusions}

In this manuscript we argued that the $DK$ interaction is attractive
enough as to generate $DK$, $DDK$ and $DDDK$ bound states.
For this we began by assuming that the $D_{s0}^*(2317)$ is a $DK$ molecule,
which determines in turn the $DK$ interaction.
Then, by means of the Gaussian Expansion Method~\cite{Kamimura:1988zz,Hiyama:2003cu} (a method for
few-body calculations),
we have addressed the question of whether one can build up multi-component
molecular states, similar to the formation of atomic nuclei
from clusters of nucleons bound by
the nucleon-nucleon interaction.
The answer is yes. We find a bound $DDK$ trimer and a $DDDK$ tetramer.
The prediction of this trimer confirms the previous calculations
of Refs.~\cite{SanchezSanchez:2017xtl,MartinezTorres:2018zbl},
while the prediction of the tetramer is novel
to the present work.

We have checked the robustness of these predictions
against a series of uncertainties.
While the $DK$ interaction is well constrained by the existence of
the $D_{s0}^*(2317)$ and chiral perturbation theory,
the $DD$ interaction is considerably
less well-known.
Yet it also enters the calculations. We chose to describe the $DD$ potential
in terms of the OBE model, in which the $DD$ interaction turns out
to be mildly attractive and has a minor impact on the binding
energy of the trimer and tetramer states.
The $DK$ potential, though well-known, is still subject to subleading
corrections, which we take into account by varying
the exact form of this potential.
As expected from the fact that we are dealing with subleading corrections,
the predictions are almost left unchanged by these variations.

In addition, we have studied a rather unlikely scenario that the $D_{s0}^*(2317)$ is dominanty a genuine $c\bar{s}$
state. Nonetheless, even in such a case, we still predict $D D_{s0}^*$ and $DDD_{s0}^*$ bound states
with the same quantum numbers as the $DDK$ trimer and $DDDK$ tetramer,
but this time located at approximately $(50-62)$ and $(60-90)\,{\rm MeV}$
below the $DDK$ and $DDDK$ thresholds (instead of $70$ and $100\,{\rm MeV}$
when the $D_{s0}^*$ is a molecular meson).
The binding mechanism is the long-range one-kaon-exchange potential
in the $D D_{s0}^*$ system: owing to the mass difference between the
$D$ and $D_{s0}^*$ mesons, the kaon is exchanged near the mass shell,
leading to an enhancement in the range of
the potential~\cite{SanchezSanchez:2017xtl}.

Although the existence of the $DDK$ and $DDDK$ bound states
seems to be quite robust, the question of where to find
them is much more challenging.
If we now focus on the $DDK$ state, the experimental discovery of
the $D_{s0}^*(2317)$ gives a clue.
As already argued in Ref.~\cite{MartinezTorres:2018zbl},
but awaiting for a concrete study, the $DDK$ state
can decay into $D D_s^*$ or $D^* D_s$ in P-wave.
Therefore one may look for inclusive combinations of three particles
$D D_s \pi$ and search for structures in the corresponding invariant
mass distributions.
Given enough statistics, there should be a possibility to discover it
in the $e^+ e^-$ collision data collected by Belle or BelleII or
in the $pp$ collision data collected at the LHC. 

It is well known that heavy quark spin and flavor symmetries relate
the $DK$ interaction to those of $D^*K$, $B\bar{K}$ and $B^*\bar{K}$.
This is consistent with the existence of the $D_{s1}(2460)$.
The bottom counterparts of the $D_{s0}^*(2317)$ and $D_{s1}(2460)$
have been predicted in a number of studies~\cite{Guo:2006fu,Guo:2006rp,Altenbuchinger:2013vwa}
and confirmed by lattice QCD simulations~\cite{Lang:2015hza}.
As a result, we naively expect the existence of the heavy quark
symmetry partners of the $DDK$ and $DDDK$ states.
At this moment, given the accessible center of mass energies at current
facilities, and the simplification that both the $D$ and $K$ 
are $0^-$ mesons that only decay weakly, we believe that they should
be of top priority both experimentally and theoretically.

\section{Acknowledgements}
This work is partly supported by the National Natural Science Foundation
of China under Grant No. 11735003, the Fundamental Research Funds
for the Central Universities, and the Thousand Talents Plan
for Young Professionals.

\bibliography{DDDK.bib}

\begin{thebibliography}{65}%
\makeatletter
\providecommand \@ifxundefined [1]{%
 \@ifx{#1\undefined}
}%
\providecommand \@ifnum [1]{%
 \ifnum #1\expandafter \@firstoftwo
 \else \expandafter \@secondoftwo
 \fi
}%
\providecommand \@ifx [1]{%
 \ifx #1\expandafter \@firstoftwo
 \else \expandafter \@secondoftwo
 \fi
}%
\providecommand \natexlab [1]{#1}%
\providecommand \enquote  [1]{``#1''}%
\providecommand \bibnamefont  [1]{#1}%
\providecommand \bibfnamefont [1]{#1}%
\providecommand \citenamefont [1]{#1}%
\providecommand \href@noop [0]{\@secondoftwo}%
\providecommand \href [0]{\begingroup \@sanitize@url \@href}%
\providecommand \@href[1]{\@@startlink{#1}\@@href}%
\providecommand \@@href[1]{\endgroup#1\@@endlink}%
\providecommand \@sanitize@url [0]{\catcode `\\12\catcode `\$12\catcode
  `\&12\catcode `\#12\catcode `\^12\catcode `\_12\catcode `\%12\relax}%
\providecommand \@@startlink[1]{}%
\providecommand \@@endlink[0]{}%
\providecommand \url  [0]{\begingroup\@sanitize@url \@url }%
\providecommand \@url [1]{\endgroup\@href {#1}{\urlprefix }}%
\providecommand \urlprefix  [0]{URL }%
\providecommand \Eprint [0]{\href }%
\providecommand \doibase [0]{http://dx.doi.org/}%
\providecommand \selectlanguage [0]{\@gobble}%
\providecommand \bibinfo  [0]{\@secondoftwo}%
\providecommand \bibfield  [0]{\@secondoftwo}%
\providecommand \translation [1]{[#1]}%
\providecommand \BibitemOpen [0]{}%
\providecommand \bibitemStop [0]{}%
\providecommand \bibitemNoStop [0]{.\EOS\space}%
\providecommand \EOS [0]{\spacefactor3000\relax}%
\providecommand \BibitemShut  [1]{\csname bibitem#1\endcsname}%
\let\auto@bib@innerbib\@empty
\bibitem [{\citenamefont {Aubert}\ \emph {et~al.}(2003)\citenamefont {Aubert}
  \emph {et~al.}}]{Aubert:2003fg}%
  \BibitemOpen
  \bibfield  {author} {\bibinfo {author} {\bibfnamefont {B.}~\bibnamefont
  {Aubert}} \emph {et~al.} (\bibinfo {collaboration} {BaBar}),\ }\href
  {\doibase 10.1103/PhysRevLett.90.242001} {\bibfield  {journal} {\bibinfo
  {journal} {Phys. Rev. Lett.}\ }\textbf {\bibinfo {volume} {90}},\ \bibinfo
  {pages} {242001} (\bibinfo {year} {2003})},\ \Eprint
  {http://arxiv.org/abs/hep-ex/0304021} {arXiv:hep-ex/0304021 [hep-ex]}
  \BibitemShut {NoStop}%
\bibitem [{\citenamefont {Besson}\ \emph {et~al.}(2003)\citenamefont {Besson}
  \emph {et~al.}}]{Besson:2003cp}%
  \BibitemOpen
  \bibfield  {author} {\bibinfo {author} {\bibfnamefont {D.}~\bibnamefont
  {Besson}} \emph {et~al.} (\bibinfo {collaboration} {CLEO}),\ }\href {\doibase
  10.1103/PhysRevD.68.032002, 10.1103/PhysRevD.75.119908} {\bibfield  {journal}
  {\bibinfo  {journal} {Phys. Rev.}\ }\textbf {\bibinfo {volume} {D68}},\
  \bibinfo {pages} {032002} (\bibinfo {year} {2003})},\ \bibinfo {note}
  {[Erratum: Phys. Rev.D75,119908(2007)]},\ \Eprint
  {http://arxiv.org/abs/hep-ex/0305100} {arXiv:hep-ex/0305100 [hep-ex]}
  \BibitemShut {NoStop}%
\bibitem [{\citenamefont {Krokovny}\ \emph {et~al.}(2003)\citenamefont
  {Krokovny} \emph {et~al.}}]{Krokovny:2003zq}%
  \BibitemOpen
  \bibfield  {author} {\bibinfo {author} {\bibfnamefont {P.}~\bibnamefont
  {Krokovny}} \emph {et~al.} (\bibinfo {collaboration} {Belle}),\ }\href
  {\doibase 10.1103/PhysRevLett.91.262002} {\bibfield  {journal} {\bibinfo
  {journal} {Phys. Rev. Lett.}\ }\textbf {\bibinfo {volume} {91}},\ \bibinfo
  {pages} {262002} (\bibinfo {year} {2003})},\ \Eprint
  {http://arxiv.org/abs/hep-ex/0308019} {arXiv:hep-ex/0308019 [hep-ex]}
  \BibitemShut {NoStop}%
\bibitem [{\citenamefont {Bardeen}\ \emph {et~al.}(2003)\citenamefont
  {Bardeen}, \citenamefont {Eichten},\ and\ \citenamefont
  {Hill}}]{Bardeen:2003kt}%
  \BibitemOpen
  \bibfield  {author} {\bibinfo {author} {\bibfnamefont {W.~A.}\ \bibnamefont
  {Bardeen}}, \bibinfo {author} {\bibfnamefont {E.~J.}\ \bibnamefont
  {Eichten}}, \ and\ \bibinfo {author} {\bibfnamefont {C.~T.}\ \bibnamefont
  {Hill}},\ }\href {\doibase 10.1103/PhysRevD.68.054024} {\bibfield  {journal}
  {\bibinfo  {journal} {Phys. Rev.}\ }\textbf {\bibinfo {volume} {D68}},\
  \bibinfo {pages} {054024} (\bibinfo {year} {2003})},\ \Eprint
  {http://arxiv.org/abs/hep-ph/0305049} {arXiv:hep-ph/0305049 [hep-ph]}
  \BibitemShut {NoStop}%
\bibitem [{\citenamefont {Nowak}\ \emph {et~al.}(2004)\citenamefont {Nowak},
  \citenamefont {Rho},\ and\ \citenamefont {Zahed}}]{Nowak:2003ra}%
  \BibitemOpen
  \bibfield  {author} {\bibinfo {author} {\bibfnamefont {M.~A.}\ \bibnamefont
  {Nowak}}, \bibinfo {author} {\bibfnamefont {M.}~\bibnamefont {Rho}}, \ and\
  \bibinfo {author} {\bibfnamefont {I.}~\bibnamefont {Zahed}},\ }\href@noop {}
  {\bibfield  {journal} {\bibinfo  {journal} {Acta Phys. Polon.}\ }\textbf
  {\bibinfo {volume} {B35}},\ \bibinfo {pages} {2377} (\bibinfo {year}
  {2004})},\ \Eprint {http://arxiv.org/abs/hep-ph/0307102}
  {arXiv:hep-ph/0307102 [hep-ph]} \BibitemShut {NoStop}%
\bibitem [{\citenamefont {van Beveren}\ and\ \citenamefont
  {Rupp}(2003)}]{vanBeveren:2003kd}%
  \BibitemOpen
  \bibfield  {author} {\bibinfo {author} {\bibfnamefont {E.}~\bibnamefont {van
  Beveren}}\ and\ \bibinfo {author} {\bibfnamefont {G.}~\bibnamefont {Rupp}},\
  }\href {\doibase 10.1103/PhysRevLett.91.012003} {\bibfield  {journal}
  {\bibinfo  {journal} {Phys. Rev. Lett.}\ }\textbf {\bibinfo {volume} {91}},\
  \bibinfo {pages} {012003} (\bibinfo {year} {2003})},\ \Eprint
  {http://arxiv.org/abs/hep-ph/0305035} {arXiv:hep-ph/0305035 [hep-ph]}
  \BibitemShut {NoStop}%
\bibitem [{\citenamefont {Dai}\ \emph {et~al.}(2003)\citenamefont {Dai},
  \citenamefont {Huang}, \citenamefont {Liu},\ and\ \citenamefont
  {Zhu}}]{Dai:2003yg}%
  \BibitemOpen
  \bibfield  {author} {\bibinfo {author} {\bibfnamefont {Y.-B.}\ \bibnamefont
  {Dai}}, \bibinfo {author} {\bibfnamefont {C.-S.}\ \bibnamefont {Huang}},
  \bibinfo {author} {\bibfnamefont {C.}~\bibnamefont {Liu}}, \ and\ \bibinfo
  {author} {\bibfnamefont {S.-L.}\ \bibnamefont {Zhu}},\ }\href {\doibase
  10.1103/PhysRevD.68.114011} {\bibfield  {journal} {\bibinfo  {journal} {Phys.
  Rev.}\ }\textbf {\bibinfo {volume} {D68}},\ \bibinfo {pages} {114011}
  (\bibinfo {year} {2003})},\ \Eprint {http://arxiv.org/abs/hep-ph/0306274}
  {arXiv:hep-ph/0306274 [hep-ph]} \BibitemShut {NoStop}%
\bibitem [{\citenamefont {Narison}(2005)}]{Narison:2003td}%
  \BibitemOpen
  \bibfield  {author} {\bibinfo {author} {\bibfnamefont {S.}~\bibnamefont
  {Narison}},\ }\href {\doibase 10.1016/j.physletb.2004.11.002} {\bibfield
  {journal} {\bibinfo  {journal} {Phys. Lett.}\ }\textbf {\bibinfo {volume}
  {B605}},\ \bibinfo {pages} {319} (\bibinfo {year} {2005})},\ \Eprint
  {http://arxiv.org/abs/hep-ph/0307248} {arXiv:hep-ph/0307248 [hep-ph]}
  \BibitemShut {NoStop}%
\bibitem [{\citenamefont {Szczepaniak}(2003)}]{Szczepaniak:2003vy}%
  \BibitemOpen
  \bibfield  {author} {\bibinfo {author} {\bibfnamefont {A.~P.}\ \bibnamefont
  {Szczepaniak}},\ }\href {\doibase 10.1016/S0370-2693(03)00865-7} {\bibfield
  {journal} {\bibinfo  {journal} {Phys. Lett.}\ }\textbf {\bibinfo {volume}
  {B567}},\ \bibinfo {pages} {23} (\bibinfo {year} {2003})},\ \Eprint
  {http://arxiv.org/abs/hep-ph/0305060} {arXiv:hep-ph/0305060 [hep-ph]}
  \BibitemShut {NoStop}%
\bibitem [{\citenamefont {Browder}\ \emph {et~al.}(2004)\citenamefont
  {Browder}, \citenamefont {Pakvasa},\ and\ \citenamefont
  {Petrov}}]{Browder:2003fk}%
  \BibitemOpen
  \bibfield  {author} {\bibinfo {author} {\bibfnamefont {T.~E.}\ \bibnamefont
  {Browder}}, \bibinfo {author} {\bibfnamefont {S.}~\bibnamefont {Pakvasa}}, \
  and\ \bibinfo {author} {\bibfnamefont {A.~A.}\ \bibnamefont {Petrov}},\
  }\href {\doibase 10.1016/j.physletb.2003.10.067} {\bibfield  {journal}
  {\bibinfo  {journal} {Phys. Lett.}\ }\textbf {\bibinfo {volume} {B578}},\
  \bibinfo {pages} {365} (\bibinfo {year} {2004})},\ \Eprint
  {http://arxiv.org/abs/hep-ph/0307054} {arXiv:hep-ph/0307054 [hep-ph]}
  \BibitemShut {NoStop}%
\bibitem [{\citenamefont {Barnes}\ \emph {et~al.}(2003)\citenamefont {Barnes},
  \citenamefont {Close},\ and\ \citenamefont {Lipkin}}]{Barnes:2003dj}%
  \BibitemOpen
  \bibfield  {author} {\bibinfo {author} {\bibfnamefont {T.}~\bibnamefont
  {Barnes}}, \bibinfo {author} {\bibfnamefont {F.~E.}\ \bibnamefont {Close}}, \
  and\ \bibinfo {author} {\bibfnamefont {H.~J.}\ \bibnamefont {Lipkin}},\
  }\href {\doibase 10.1103/PhysRevD.68.054006} {\bibfield  {journal} {\bibinfo
  {journal} {Phys. Rev.}\ }\textbf {\bibinfo {volume} {D68}},\ \bibinfo {pages}
  {054006} (\bibinfo {year} {2003})},\ \Eprint
  {http://arxiv.org/abs/hep-ph/0305025} {arXiv:hep-ph/0305025 [hep-ph]}
  \BibitemShut {NoStop}%
\bibitem [{\citenamefont {Cheng}\ and\ \citenamefont
  {Hou}(2003)}]{Cheng:2003kg}%
  \BibitemOpen
  \bibfield  {author} {\bibinfo {author} {\bibfnamefont {H.-Y.}\ \bibnamefont
  {Cheng}}\ and\ \bibinfo {author} {\bibfnamefont {W.-S.}\ \bibnamefont
  {Hou}},\ }\href {\doibase 10.1016/S0370-2693(03)00834-7} {\bibfield
  {journal} {\bibinfo  {journal} {Phys. Lett.}\ }\textbf {\bibinfo {volume}
  {B566}},\ \bibinfo {pages} {193} (\bibinfo {year} {2003})},\ \Eprint
  {http://arxiv.org/abs/hep-ph/0305038} {arXiv:hep-ph/0305038 [hep-ph]}
  \BibitemShut {NoStop}%
\bibitem [{\citenamefont {Chen}\ and\ \citenamefont {Li}(2004)}]{Chen:2004dy}%
  \BibitemOpen
  \bibfield  {author} {\bibinfo {author} {\bibfnamefont {Y.-Q.}\ \bibnamefont
  {Chen}}\ and\ \bibinfo {author} {\bibfnamefont {X.-Q.}\ \bibnamefont {Li}},\
  }\href {\doibase 10.1103/PhysRevLett.93.232001} {\bibfield  {journal}
  {\bibinfo  {journal} {Phys. Rev. Lett.}\ }\textbf {\bibinfo {volume} {93}},\
  \bibinfo {pages} {232001} (\bibinfo {year} {2004})},\ \Eprint
  {http://arxiv.org/abs/hep-ph/0407062} {arXiv:hep-ph/0407062 [hep-ph]}
  \BibitemShut {NoStop}%
\bibitem [{\citenamefont {Dmitrasinovic}(2005)}]{Dmitrasinovic:2005gc}%
  \BibitemOpen
  \bibfield  {author} {\bibinfo {author} {\bibfnamefont {V.}~\bibnamefont
  {Dmitrasinovic}},\ }\href {\doibase 10.1103/PhysRevLett.94.162002} {\bibfield
   {journal} {\bibinfo  {journal} {Phys. Rev. Lett.}\ }\textbf {\bibinfo
  {volume} {94}},\ \bibinfo {pages} {162002} (\bibinfo {year}
  {2005})}\BibitemShut {NoStop}%
\bibitem [{\citenamefont {Zhang}(2019)}]{Zhang:2018mnm}%
  \BibitemOpen
  \bibfield  {author} {\bibinfo {author} {\bibfnamefont {J.-R.}\ \bibnamefont
  {Zhang}},\ }\href {\doibase 10.1016/j.physletb.2019.01.001} {\bibfield
  {journal} {\bibinfo  {journal} {Phys. Lett.}\ }\textbf {\bibinfo {volume}
  {B789}},\ \bibinfo {pages} {432} (\bibinfo {year} {2019})},\ \Eprint
  {http://arxiv.org/abs/1801.08725} {arXiv:1801.08725 [hep-ph]} \BibitemShut
  {NoStop}%
\bibitem [{\citenamefont {Terasaki}(2003)}]{Terasaki:2003qa}%
  \BibitemOpen
  \bibfield  {author} {\bibinfo {author} {\bibfnamefont {K.}~\bibnamefont
  {Terasaki}},\ }\href {\doibase 10.1103/PhysRevD.68.011501} {\bibfield
  {journal} {\bibinfo  {journal} {Phys. Rev.}\ }\textbf {\bibinfo {volume}
  {D68}},\ \bibinfo {pages} {011501} (\bibinfo {year} {2003})},\ \Eprint
  {http://arxiv.org/abs/hep-ph/0305213} {arXiv:hep-ph/0305213 [hep-ph]}
  \BibitemShut {NoStop}%
\bibitem [{\citenamefont {Maiani}\ \emph {et~al.}(2005)\citenamefont {Maiani},
  \citenamefont {Piccinini}, \citenamefont {Polosa},\ and\ \citenamefont
  {Riquer}}]{Maiani:2004vq}%
  \BibitemOpen
  \bibfield  {author} {\bibinfo {author} {\bibfnamefont {L.}~\bibnamefont
  {Maiani}}, \bibinfo {author} {\bibfnamefont {F.}~\bibnamefont {Piccinini}},
  \bibinfo {author} {\bibfnamefont {A.}~\bibnamefont {Polosa}}, \ and\ \bibinfo
  {author} {\bibfnamefont {V.}~\bibnamefont {Riquer}},\ }\href {\doibase
  10.1103/PhysRevD.71.014028} {\bibfield  {journal} {\bibinfo  {journal}
  {Phys.Rev.}\ }\textbf {\bibinfo {volume} {D71}},\ \bibinfo {pages} {014028}
  (\bibinfo {year} {2005})},\ \Eprint {http://arxiv.org/abs/hep-ph/0412098}
  {arXiv:hep-ph/0412098 [hep-ph]} \BibitemShut {NoStop}%
\bibitem [{\citenamefont {Kolomeitsev}\ and\ \citenamefont
  {Lutz}(2004)}]{Kolomeitsev:2003ac}%
  \BibitemOpen
  \bibfield  {author} {\bibinfo {author} {\bibfnamefont {E.~E.}\ \bibnamefont
  {Kolomeitsev}}\ and\ \bibinfo {author} {\bibfnamefont {M.~F.~M.}\
  \bibnamefont {Lutz}},\ }\href {\doibase 10.1016/j.physletb.2003.10.118}
  {\bibfield  {journal} {\bibinfo  {journal} {Phys. Lett.}\ }\textbf {\bibinfo
  {volume} {B582}},\ \bibinfo {pages} {39} (\bibinfo {year} {2004})},\ \Eprint
  {http://arxiv.org/abs/hep-ph/0307133} {arXiv:hep-ph/0307133 [hep-ph]}
  \BibitemShut {NoStop}%
\bibitem [{\citenamefont {Hofmann}\ and\ \citenamefont
  {Lutz}(2004)}]{Hofmann:2003je}%
  \BibitemOpen
  \bibfield  {author} {\bibinfo {author} {\bibfnamefont {J.}~\bibnamefont
  {Hofmann}}\ and\ \bibinfo {author} {\bibfnamefont {M.~F.~M.}\ \bibnamefont
  {Lutz}},\ }\href {\doibase 10.1016/j.nuclphysa.2003.12.013} {\bibfield
  {journal} {\bibinfo  {journal} {Nucl. Phys.}\ }\textbf {\bibinfo {volume}
  {A733}},\ \bibinfo {pages} {142} (\bibinfo {year} {2004})},\ \Eprint
  {http://arxiv.org/abs/hep-ph/0308263} {arXiv:hep-ph/0308263 [hep-ph]}
  \BibitemShut {NoStop}%
\bibitem [{\citenamefont {Guo}\ \emph {et~al.}(2008)\citenamefont {Guo},
  \citenamefont {Hanhart}, \citenamefont {Krewald},\ and\ \citenamefont
  {Meissner}}]{Guo:2008gp}%
  \BibitemOpen
  \bibfield  {author} {\bibinfo {author} {\bibfnamefont {F.-K.}\ \bibnamefont
  {Guo}}, \bibinfo {author} {\bibfnamefont {C.}~\bibnamefont {Hanhart}},
  \bibinfo {author} {\bibfnamefont {S.}~\bibnamefont {Krewald}}, \ and\
  \bibinfo {author} {\bibfnamefont {U.-G.}\ \bibnamefont {Meissner}},\ }\href
  {\doibase 10.1016/j.physletb.2008.07.060} {\bibfield  {journal} {\bibinfo
  {journal} {Phys. Lett.}\ }\textbf {\bibinfo {volume} {B666}},\ \bibinfo
  {pages} {251} (\bibinfo {year} {2008})},\ \Eprint
  {http://arxiv.org/abs/0806.3374} {arXiv:0806.3374 [hep-ph]} \BibitemShut
  {NoStop}%
\bibitem [{\citenamefont {Guo}\ \emph {et~al.}(2006)\citenamefont {Guo},
  \citenamefont {Shen}, \citenamefont {Chiang}, \citenamefont {Ping},\ and\
  \citenamefont {Zou}}]{Guo:2006fu}%
  \BibitemOpen
  \bibfield  {author} {\bibinfo {author} {\bibfnamefont {F.-K.}\ \bibnamefont
  {Guo}}, \bibinfo {author} {\bibfnamefont {P.-N.}\ \bibnamefont {Shen}},
  \bibinfo {author} {\bibfnamefont {H.-C.}\ \bibnamefont {Chiang}}, \bibinfo
  {author} {\bibfnamefont {R.-G.}\ \bibnamefont {Ping}}, \ and\ \bibinfo
  {author} {\bibfnamefont {B.-S.}\ \bibnamefont {Zou}},\ }\href {\doibase
  10.1016/j.physletb.2006.08.064} {\bibfield  {journal} {\bibinfo  {journal}
  {Phys. Lett.}\ }\textbf {\bibinfo {volume} {B641}},\ \bibinfo {pages} {278}
  (\bibinfo {year} {2006})},\ \Eprint {http://arxiv.org/abs/hep-ph/0603072}
  {arXiv:hep-ph/0603072 [hep-ph]} \BibitemShut {NoStop}%
\bibitem [{\citenamefont {Guo}\ \emph {et~al.}(2009)\citenamefont {Guo},
  \citenamefont {Hanhart},\ and\ \citenamefont {Meissner}}]{Guo:2009ct}%
  \BibitemOpen
  \bibfield  {author} {\bibinfo {author} {\bibfnamefont {F.-K.}\ \bibnamefont
  {Guo}}, \bibinfo {author} {\bibfnamefont {C.}~\bibnamefont {Hanhart}}, \ and\
  \bibinfo {author} {\bibfnamefont {U.-G.}\ \bibnamefont {Meissner}},\ }\href
  {\doibase 10.1140/epja/i2009-10762-1} {\bibfield  {journal} {\bibinfo
  {journal} {Eur. Phys. J.}\ }\textbf {\bibinfo {volume} {A40}},\ \bibinfo
  {pages} {171} (\bibinfo {year} {2009})},\ \Eprint
  {http://arxiv.org/abs/0901.1597} {arXiv:0901.1597 [hep-ph]} \BibitemShut
  {NoStop}%
\bibitem [{\citenamefont {Cleven}\ \emph {et~al.}(2011)\citenamefont {Cleven},
  \citenamefont {Guo}, \citenamefont {Hanhart},\ and\ \citenamefont
  {Meissner}}]{Cleven:2010aw}%
  \BibitemOpen
  \bibfield  {author} {\bibinfo {author} {\bibfnamefont {M.}~\bibnamefont
  {Cleven}}, \bibinfo {author} {\bibfnamefont {F.-K.}\ \bibnamefont {Guo}},
  \bibinfo {author} {\bibfnamefont {C.}~\bibnamefont {Hanhart}}, \ and\
  \bibinfo {author} {\bibfnamefont {U.-G.}\ \bibnamefont {Meissner}},\ }\href
  {\doibase 10.1140/epja/i2011-11019-2} {\bibfield  {journal} {\bibinfo
  {journal} {Eur. Phys. J.}\ }\textbf {\bibinfo {volume} {A47}},\ \bibinfo
  {pages} {19} (\bibinfo {year} {2011})},\ \Eprint
  {http://arxiv.org/abs/1009.3804} {arXiv:1009.3804 [hep-ph]} \BibitemShut
  {NoStop}%
\bibitem [{\citenamefont {Martinez~Torres}\ \emph {et~al.}(2012)\citenamefont
  {Martinez~Torres}, \citenamefont {Dai}, \citenamefont {Koren}, \citenamefont
  {Jido},\ and\ \citenamefont {Oset}}]{MartinezTorres:2011pr}%
  \BibitemOpen
  \bibfield  {author} {\bibinfo {author} {\bibfnamefont {A.}~\bibnamefont
  {Martinez~Torres}}, \bibinfo {author} {\bibfnamefont {L.~R.}\ \bibnamefont
  {Dai}}, \bibinfo {author} {\bibfnamefont {C.}~\bibnamefont {Koren}}, \bibinfo
  {author} {\bibfnamefont {D.}~\bibnamefont {Jido}}, \ and\ \bibinfo {author}
  {\bibfnamefont {E.}~\bibnamefont {Oset}},\ }\href {\doibase
  10.1103/PhysRevD.85.014027} {\bibfield  {journal} {\bibinfo  {journal} {Phys.
  Rev.}\ }\textbf {\bibinfo {volume} {D85}},\ \bibinfo {pages} {014027}
  (\bibinfo {year} {2012})},\ \Eprint {http://arxiv.org/abs/1109.0396}
  {arXiv:1109.0396 [hep-lat]} \BibitemShut {NoStop}%
\bibitem [{\citenamefont {Martínez~Torres}\ \emph {et~al.}(2015)\citenamefont
  {Martínez~Torres}, \citenamefont {Oset}, \citenamefont {Prelovsek},\ and\
  \citenamefont {Ramos}}]{Torres:2014vna}%
  \BibitemOpen
  \bibfield  {author} {\bibinfo {author} {\bibfnamefont {A.}~\bibnamefont
  {Martínez~Torres}}, \bibinfo {author} {\bibfnamefont {E.}~\bibnamefont
  {Oset}}, \bibinfo {author} {\bibfnamefont {S.}~\bibnamefont {Prelovsek}}, \
  and\ \bibinfo {author} {\bibfnamefont {A.}~\bibnamefont {Ramos}},\ }\href
  {\doibase 10.1007/JHEP05(2015)153} {\bibfield  {journal} {\bibinfo  {journal}
  {JHEP}\ }\textbf {\bibinfo {volume} {05}},\ \bibinfo {pages} {153} (\bibinfo
  {year} {2015})},\ \Eprint {http://arxiv.org/abs/1412.1706} {arXiv:1412.1706
  [hep-lat]} \BibitemShut {NoStop}%
\bibitem [{\citenamefont {Yao}\ \emph {et~al.}(2015)\citenamefont {Yao},
  \citenamefont {Du}, \citenamefont {Guo},\ and\ \citenamefont
  {Meißner}}]{Yao:2015qia}%
  \BibitemOpen
  \bibfield  {author} {\bibinfo {author} {\bibfnamefont {D.-L.}\ \bibnamefont
  {Yao}}, \bibinfo {author} {\bibfnamefont {M.-L.}\ \bibnamefont {Du}},
  \bibinfo {author} {\bibfnamefont {F.-K.}\ \bibnamefont {Guo}}, \ and\
  \bibinfo {author} {\bibfnamefont {U.-G.}\ \bibnamefont {Meißner}},\ }\href
  {\doibase 10.1007/JHEP11(2015)058} {\bibfield  {journal} {\bibinfo  {journal}
  {JHEP}\ }\textbf {\bibinfo {volume} {11}},\ \bibinfo {pages} {058} (\bibinfo
  {year} {2015})},\ \Eprint {http://arxiv.org/abs/1502.05981} {arXiv:1502.05981
  [hep-ph]} \BibitemShut {NoStop}%
\bibitem [{\citenamefont {Guo}\ \emph {et~al.}(2015)\citenamefont {Guo},
  \citenamefont {Meißner},\ and\ \citenamefont {Yao}}]{Guo:2015dha}%
  \BibitemOpen
  \bibfield  {author} {\bibinfo {author} {\bibfnamefont {Z.-H.}\ \bibnamefont
  {Guo}}, \bibinfo {author} {\bibfnamefont {U.-G.}\ \bibnamefont {Meißner}}, \
  and\ \bibinfo {author} {\bibfnamefont {D.-L.}\ \bibnamefont {Yao}},\ }\href
  {\doibase 10.1103/PhysRevD.92.094008} {\bibfield  {journal} {\bibinfo
  {journal} {Phys. Rev.}\ }\textbf {\bibinfo {volume} {D92}},\ \bibinfo {pages}
  {094008} (\bibinfo {year} {2015})},\ \Eprint
  {http://arxiv.org/abs/1507.03123} {arXiv:1507.03123 [hep-ph]} \BibitemShut
  {NoStop}%
\bibitem [{\citenamefont {Albaladejo}\ \emph {et~al.}(2017)\citenamefont
  {Albaladejo}, \citenamefont {Fernandez-Soler}, \citenamefont {Guo},\ and\
  \citenamefont {Nieves}}]{Albaladejo:2016lbb}%
  \BibitemOpen
  \bibfield  {author} {\bibinfo {author} {\bibfnamefont {M.}~\bibnamefont
  {Albaladejo}}, \bibinfo {author} {\bibfnamefont {P.}~\bibnamefont
  {Fernandez-Soler}}, \bibinfo {author} {\bibfnamefont {F.-K.}\ \bibnamefont
  {Guo}}, \ and\ \bibinfo {author} {\bibfnamefont {J.}~\bibnamefont {Nieves}},\
  }\href {\doibase 10.1016/j.physletb.2017.02.036} {\bibfield  {journal}
  {\bibinfo  {journal} {Phys. Lett.}\ }\textbf {\bibinfo {volume} {B767}},\
  \bibinfo {pages} {465} (\bibinfo {year} {2017})},\ \Eprint
  {http://arxiv.org/abs/1610.06727} {arXiv:1610.06727 [hep-ph]} \BibitemShut
  {NoStop}%
\bibitem [{\citenamefont {Du}\ \emph {et~al.}(2017)\citenamefont {Du},
  \citenamefont {Guo}, \citenamefont {Meißner},\ and\ \citenamefont
  {Yao}}]{Du:2017ttu}%
  \BibitemOpen
  \bibfield  {author} {\bibinfo {author} {\bibfnamefont {M.-L.}\ \bibnamefont
  {Du}}, \bibinfo {author} {\bibfnamefont {F.-K.}\ \bibnamefont {Guo}},
  \bibinfo {author} {\bibfnamefont {U.-G.}\ \bibnamefont {Meißner}}, \ and\
  \bibinfo {author} {\bibfnamefont {D.-L.}\ \bibnamefont {Yao}},\ }\href
  {\doibase 10.1140/epjc/s10052-017-5287-6} {\bibfield  {journal} {\bibinfo
  {journal} {Eur. Phys. J.}\ }\textbf {\bibinfo {volume} {C77}},\ \bibinfo
  {pages} {728} (\bibinfo {year} {2017})},\ \Eprint
  {http://arxiv.org/abs/1703.10836} {arXiv:1703.10836 [hep-ph]} \BibitemShut
  {NoStop}%
\bibitem [{\citenamefont {Guo}\ \emph {et~al.}(2018{\natexlab{a}})\citenamefont
  {Guo}, \citenamefont {Heo},\ and\ \citenamefont {Lutz}}]{Guo:2018kno}%
  \BibitemOpen
  \bibfield  {author} {\bibinfo {author} {\bibfnamefont {X.-Y.}\ \bibnamefont
  {Guo}}, \bibinfo {author} {\bibfnamefont {Y.}~\bibnamefont {Heo}}, \ and\
  \bibinfo {author} {\bibfnamefont {M.~F.~M.}\ \bibnamefont {Lutz}},\ }\href
  {\doibase 10.1103/PhysRevD.98.014510} {\bibfield  {journal} {\bibinfo
  {journal} {Phys. Rev.}\ }\textbf {\bibinfo {volume} {D98}},\ \bibinfo {pages}
  {014510} (\bibinfo {year} {2018}{\natexlab{a}})},\ \Eprint
  {http://arxiv.org/abs/1801.10122} {arXiv:1801.10122 [hep-lat]} \BibitemShut
  {NoStop}%
\bibitem [{\citenamefont {Albaladejo}\ \emph {et~al.}(2018)\citenamefont
  {Albaladejo}, \citenamefont {Fernandez-Soler}, \citenamefont {Nieves},\ and\
  \citenamefont {Ortega}}]{Albaladejo:2018mhb}%
  \BibitemOpen
  \bibfield  {author} {\bibinfo {author} {\bibfnamefont {M.}~\bibnamefont
  {Albaladejo}}, \bibinfo {author} {\bibfnamefont {P.}~\bibnamefont
  {Fernandez-Soler}}, \bibinfo {author} {\bibfnamefont {J.}~\bibnamefont
  {Nieves}}, \ and\ \bibinfo {author} {\bibfnamefont {P.~G.}\ \bibnamefont
  {Ortega}},\ }\href {\doibase 10.1140/epjc/s10052-018-6176-3} {\bibfield
  {journal} {\bibinfo  {journal} {Eur. Phys. J.}\ }\textbf {\bibinfo {volume}
  {C78}},\ \bibinfo {pages} {722} (\bibinfo {year} {2018})},\ \Eprint
  {http://arxiv.org/abs/1805.07104} {arXiv:1805.07104 [hep-ph]} \BibitemShut
  {NoStop}%
\bibitem [{\citenamefont {Altenbuchinger}\ and\ \citenamefont
  {Geng}(2014)}]{Altenbuchinger:2013gaa}%
  \BibitemOpen
  \bibfield  {author} {\bibinfo {author} {\bibfnamefont {M.}~\bibnamefont
  {Altenbuchinger}}\ and\ \bibinfo {author} {\bibfnamefont {L.-S.}\
  \bibnamefont {Geng}},\ }\href {\doibase 10.1103/PhysRevD.89.054008}
  {\bibfield  {journal} {\bibinfo  {journal} {Phys. Rev.}\ }\textbf {\bibinfo
  {volume} {D89}},\ \bibinfo {pages} {054008} (\bibinfo {year} {2014})},\
  \Eprint {http://arxiv.org/abs/1310.5224} {arXiv:1310.5224 [hep-ph]}
  \BibitemShut {NoStop}%
\bibitem [{\citenamefont {Altenbuchinger}\ \emph {et~al.}(2014)\citenamefont
  {Altenbuchinger}, \citenamefont {Geng},\ and\ \citenamefont
  {Weise}}]{Altenbuchinger:2013vwa}%
  \BibitemOpen
  \bibfield  {author} {\bibinfo {author} {\bibfnamefont {M.}~\bibnamefont
  {Altenbuchinger}}, \bibinfo {author} {\bibfnamefont {L.~S.}\ \bibnamefont
  {Geng}}, \ and\ \bibinfo {author} {\bibfnamefont {W.}~\bibnamefont {Weise}},\
  }\href {\doibase 10.1103/PhysRevD.89.014026} {\bibfield  {journal} {\bibinfo
  {journal} {Phys. Rev.}\ }\textbf {\bibinfo {volume} {D89}},\ \bibinfo {pages}
  {014026} (\bibinfo {year} {2014})},\ \Eprint {http://arxiv.org/abs/1309.4743}
  {arXiv:1309.4743 [hep-ph]} \BibitemShut {NoStop}%
\bibitem [{\citenamefont {Geng}\ \emph {et~al.}(2010)\citenamefont {Geng},
  \citenamefont {Kaiser}, \citenamefont {Martin-Camalich},\ and\ \citenamefont
  {Weise}}]{Geng:2010vw}%
  \BibitemOpen
  \bibfield  {author} {\bibinfo {author} {\bibfnamefont {L.~S.}\ \bibnamefont
  {Geng}}, \bibinfo {author} {\bibfnamefont {N.}~\bibnamefont {Kaiser}},
  \bibinfo {author} {\bibfnamefont {J.}~\bibnamefont {Martin-Camalich}}, \ and\
  \bibinfo {author} {\bibfnamefont {W.}~\bibnamefont {Weise}},\ }\href
  {\doibase 10.1103/PhysRevD.82.054022} {\bibfield  {journal} {\bibinfo
  {journal} {Phys. Rev.}\ }\textbf {\bibinfo {volume} {D82}},\ \bibinfo {pages}
  {054022} (\bibinfo {year} {2010})},\ \Eprint {http://arxiv.org/abs/1008.0383}
  {arXiv:1008.0383 [hep-ph]} \BibitemShut {NoStop}%
\bibitem [{\citenamefont {Wang}\ and\ \citenamefont
  {Wang}(2012)}]{Wang:2012bu}%
  \BibitemOpen
  \bibfield  {author} {\bibinfo {author} {\bibfnamefont {P.}~\bibnamefont
  {Wang}}\ and\ \bibinfo {author} {\bibfnamefont {X.~G.}\ \bibnamefont
  {Wang}},\ }\href {\doibase 10.1103/PhysRevD.86.039903,
  10.1103/PhysRevD.86.014030} {\bibfield  {journal} {\bibinfo  {journal} {Phys.
  Rev.}\ }\textbf {\bibinfo {volume} {D86}},\ \bibinfo {pages} {014030}
  (\bibinfo {year} {2012})},\ \Eprint {http://arxiv.org/abs/1204.5553}
  {arXiv:1204.5553 [hep-ph]} \BibitemShut {NoStop}%
\bibitem [{\citenamefont {Liu}\ \emph {et~al.}(2009)\citenamefont {Liu},
  \citenamefont {Liu},\ and\ \citenamefont {Zhu}}]{Liu:2009uz}%
  \BibitemOpen
  \bibfield  {author} {\bibinfo {author} {\bibfnamefont {Y.-R.}\ \bibnamefont
  {Liu}}, \bibinfo {author} {\bibfnamefont {X.}~\bibnamefont {Liu}}, \ and\
  \bibinfo {author} {\bibfnamefont {S.-L.}\ \bibnamefont {Zhu}},\ }\href
  {\doibase 10.1103/PhysRevD.79.094026} {\bibfield  {journal} {\bibinfo
  {journal} {Phys. Rev.}\ }\textbf {\bibinfo {volume} {D79}},\ \bibinfo {pages}
  {094026} (\bibinfo {year} {2009})},\ \Eprint {http://arxiv.org/abs/0904.1770}
  {arXiv:0904.1770 [hep-ph]} \BibitemShut {NoStop}%
\bibitem [{\citenamefont {Guo}\ \emph {et~al.}(2018{\natexlab{b}})\citenamefont
  {Guo}, \citenamefont {Liu}, \citenamefont {Meissner}, \citenamefont {Oller},\
  and\ \citenamefont {Rusetsky}}]{Guo:2018ocg}%
  \BibitemOpen
  \bibfield  {author} {\bibinfo {author} {\bibfnamefont {Z.-H.}\ \bibnamefont
  {Guo}}, \bibinfo {author} {\bibfnamefont {L.}~\bibnamefont {Liu}}, \bibinfo
  {author} {\bibfnamefont {U.-G.}\ \bibnamefont {Meissner}}, \bibinfo {author}
  {\bibfnamefont {J.~A.}\ \bibnamefont {Oller}}, \ and\ \bibinfo {author}
  {\bibfnamefont {A.}~\bibnamefont {Rusetsky}},\ }in\ \href@noop {} {\emph
  {\bibinfo {booktitle} {{36th International Symposium on Lattice Field Theory
  (Lattice 2018) East Lansing, MI, United States, July 22-28, 2018}}}}\
  (\bibinfo {year} {2018})\ \Eprint {http://arxiv.org/abs/1811.05582}
  {arXiv:1811.05582 [hep-lat]} \BibitemShut {NoStop}%
\bibitem [{\citenamefont {Guo}\ \emph {et~al.}(2019)\citenamefont {Guo},
  \citenamefont {Liu}, \citenamefont {Meißner}, \citenamefont {Oller},\ and\
  \citenamefont {Rusetsky}}]{Guo:2018tjx}%
  \BibitemOpen
  \bibfield  {author} {\bibinfo {author} {\bibfnamefont {Z.-H.}\ \bibnamefont
  {Guo}}, \bibinfo {author} {\bibfnamefont {L.}~\bibnamefont {Liu}}, \bibinfo
  {author} {\bibfnamefont {U.-G.}\ \bibnamefont {Meißner}}, \bibinfo {author}
  {\bibfnamefont {J.~A.}\ \bibnamefont {Oller}}, \ and\ \bibinfo {author}
  {\bibfnamefont {A.}~\bibnamefont {Rusetsky}},\ }\href {\doibase
  10.1140/epjc/s10052-018-6518-1} {\bibfield  {journal} {\bibinfo  {journal}
  {Eur. Phys. J.}\ }\textbf {\bibinfo {volume} {C79}},\ \bibinfo {pages} {13}
  (\bibinfo {year} {2019})},\ \Eprint {http://arxiv.org/abs/1811.05585}
  {arXiv:1811.05585 [hep-ph]} \BibitemShut {NoStop}%
\bibitem [{\citenamefont {Liu}\ \emph {et~al.}(2013)\citenamefont {Liu},
  \citenamefont {Orginos}, \citenamefont {Guo}, \citenamefont {Hanhart},\ and\
  \citenamefont {Meissner}}]{Liu:2012zya}%
  \BibitemOpen
  \bibfield  {author} {\bibinfo {author} {\bibfnamefont {L.}~\bibnamefont
  {Liu}}, \bibinfo {author} {\bibfnamefont {K.}~\bibnamefont {Orginos}},
  \bibinfo {author} {\bibfnamefont {F.-K.}\ \bibnamefont {Guo}}, \bibinfo
  {author} {\bibfnamefont {C.}~\bibnamefont {Hanhart}}, \ and\ \bibinfo
  {author} {\bibfnamefont {U.-G.}\ \bibnamefont {Meissner}},\ }\href {\doibase
  10.1103/PhysRevD.87.014508} {\bibfield  {journal} {\bibinfo  {journal} {Phys.
  Rev.}\ }\textbf {\bibinfo {volume} {D87}},\ \bibinfo {pages} {014508}
  (\bibinfo {year} {2013})},\ \Eprint {http://arxiv.org/abs/1208.4535}
  {arXiv:1208.4535 [hep-lat]} \BibitemShut {NoStop}%
\bibitem [{\citenamefont {Mohler}\ \emph {et~al.}(2013)\citenamefont {Mohler},
  \citenamefont {Lang}, \citenamefont {Leskovec}, \citenamefont {Prelovsek},\
  and\ \citenamefont {Woloshyn}}]{Mohler:2013rwa}%
  \BibitemOpen
  \bibfield  {author} {\bibinfo {author} {\bibfnamefont {D.}~\bibnamefont
  {Mohler}}, \bibinfo {author} {\bibfnamefont {C.~B.}\ \bibnamefont {Lang}},
  \bibinfo {author} {\bibfnamefont {L.}~\bibnamefont {Leskovec}}, \bibinfo
  {author} {\bibfnamefont {S.}~\bibnamefont {Prelovsek}}, \ and\ \bibinfo
  {author} {\bibfnamefont {R.~M.}\ \bibnamefont {Woloshyn}},\ }\href {\doibase
  10.1103/PhysRevLett.111.222001} {\bibfield  {journal} {\bibinfo  {journal}
  {Phys. Rev. Lett.}\ }\textbf {\bibinfo {volume} {111}},\ \bibinfo {pages}
  {222001} (\bibinfo {year} {2013})},\ \Eprint {http://arxiv.org/abs/1308.3175}
  {arXiv:1308.3175 [hep-lat]} \BibitemShut {NoStop}%
\bibitem [{\citenamefont {Lang}\ \emph {et~al.}(2014)\citenamefont {Lang},
  \citenamefont {Leskovec}, \citenamefont {Mohler}, \citenamefont {Prelovsek},\
  and\ \citenamefont {Woloshyn}}]{Lang:2014yfa}%
  \BibitemOpen
  \bibfield  {author} {\bibinfo {author} {\bibfnamefont {C.~B.}\ \bibnamefont
  {Lang}}, \bibinfo {author} {\bibfnamefont {L.}~\bibnamefont {Leskovec}},
  \bibinfo {author} {\bibfnamefont {D.}~\bibnamefont {Mohler}}, \bibinfo
  {author} {\bibfnamefont {S.}~\bibnamefont {Prelovsek}}, \ and\ \bibinfo
  {author} {\bibfnamefont {R.~M.}\ \bibnamefont {Woloshyn}},\ }\href {\doibase
  10.1103/PhysRevD.90.034510} {\bibfield  {journal} {\bibinfo  {journal} {Phys.
  Rev.}\ }\textbf {\bibinfo {volume} {D90}},\ \bibinfo {pages} {034510}
  (\bibinfo {year} {2014})},\ \Eprint {http://arxiv.org/abs/1403.8103}
  {arXiv:1403.8103 [hep-lat]} \BibitemShut {NoStop}%
\bibitem [{\citenamefont {Bali}\ \emph {et~al.}(2017)\citenamefont {Bali},
  \citenamefont {Collins}, \citenamefont {Cox},\ and\ \citenamefont
  {Schäfer}}]{Bali:2017pdv}%
  \BibitemOpen
  \bibfield  {author} {\bibinfo {author} {\bibfnamefont {G.~S.}\ \bibnamefont
  {Bali}}, \bibinfo {author} {\bibfnamefont {S.}~\bibnamefont {Collins}},
  \bibinfo {author} {\bibfnamefont {A.}~\bibnamefont {Cox}}, \ and\ \bibinfo
  {author} {\bibfnamefont {A.}~\bibnamefont {Schäfer}},\ }\href {\doibase
  10.1103/PhysRevD.96.074501} {\bibfield  {journal} {\bibinfo  {journal} {Phys.
  Rev.}\ }\textbf {\bibinfo {volume} {D96}},\ \bibinfo {pages} {074501}
  (\bibinfo {year} {2017})},\ \Eprint {http://arxiv.org/abs/1706.01247}
  {arXiv:1706.01247 [hep-lat]} \BibitemShut {NoStop}%
\bibitem [{\citenamefont {Guo}(2019)}]{Guo:2019dpg}%
  \BibitemOpen
  \bibfield  {author} {\bibinfo {author} {\bibfnamefont {F.-K.}\ \bibnamefont
  {Guo}},\ }\bibfield  {booktitle} {\emph {\bibinfo {booktitle} {{Proceedings,
  9th International Workshop on Charm Physics (CHARM 2018): Novosibirsk,
  Russia, May 21-25, 2018}}},\ }\href {\doibase 10.1051/epjconf/201920202001}
  {\bibfield  {journal} {\bibinfo  {journal} {EPJ Web Conf.}\ }\textbf
  {\bibinfo {volume} {202}},\ \bibinfo {pages} {02001} (\bibinfo {year}
  {2019})}\BibitemShut {NoStop}%
\bibitem [{\citenamefont {Ma}\ \emph {et~al.}(2019)\citenamefont {Ma},
  \citenamefont {Wang},\ and\ \citenamefont {Meißner}}]{Ma:2017ery}%
  \BibitemOpen
  \bibfield  {author} {\bibinfo {author} {\bibfnamefont {L.}~\bibnamefont
  {Ma}}, \bibinfo {author} {\bibfnamefont {Q.}~\bibnamefont {Wang}}, \ and\
  \bibinfo {author} {\bibfnamefont {U.-G.}\ \bibnamefont {Meißner}},\ }\href
  {\doibase 10.1088/1674-1137/43/1/014102} {\bibfield  {journal} {\bibinfo
  {journal} {Chin. Phys.}\ }\textbf {\bibinfo {volume} {C43}},\ \bibinfo
  {pages} {014102} (\bibinfo {year} {2019})},\ \Eprint
  {http://arxiv.org/abs/1711.06143} {arXiv:1711.06143 [hep-ph]} \BibitemShut
  {NoStop}%
\bibitem [{\citenamefont {Ren}\ \emph {et~al.}(2018)\citenamefont {Ren},
  \citenamefont {Malabarba}, \citenamefont {Geng}, \citenamefont
  {Khemchandani},\ and\ \citenamefont {Martínez~Torres}}]{Ren:2018pcd}%
  \BibitemOpen
  \bibfield  {author} {\bibinfo {author} {\bibfnamefont {X.-L.}\ \bibnamefont
  {Ren}}, \bibinfo {author} {\bibfnamefont {B.~B.}\ \bibnamefont {Malabarba}},
  \bibinfo {author} {\bibfnamefont {L.-S.}\ \bibnamefont {Geng}}, \bibinfo
  {author} {\bibfnamefont {K.~P.}\ \bibnamefont {Khemchandani}}, \ and\
  \bibinfo {author} {\bibfnamefont {A.}~\bibnamefont {Martínez~Torres}},\
  }\href {\doibase 10.1016/j.physletb.2018.08.034} {\bibfield  {journal}
  {\bibinfo  {journal} {Phys. Lett.}\ }\textbf {\bibinfo {volume} {B785}},\
  \bibinfo {pages} {112} (\bibinfo {year} {2018})},\ \Eprint
  {http://arxiv.org/abs/1805.08330} {arXiv:1805.08330 [hep-ph]} \BibitemShut
  {NoStop}%
\bibitem [{\citenamefont {Sanchez~Sanchez}\ \emph {et~al.}(2018)\citenamefont
  {Sanchez~Sanchez}, \citenamefont {Geng}, \citenamefont {Lu}, \citenamefont
  {Hyodo},\ and\ \citenamefont {Valderrama}}]{SanchezSanchez:2017xtl}%
  \BibitemOpen
  \bibfield  {author} {\bibinfo {author} {\bibfnamefont {M.}~\bibnamefont
  {Sanchez~Sanchez}}, \bibinfo {author} {\bibfnamefont {L.-S.}\ \bibnamefont
  {Geng}}, \bibinfo {author} {\bibfnamefont {J.-X.}\ \bibnamefont {Lu}},
  \bibinfo {author} {\bibfnamefont {T.}~\bibnamefont {Hyodo}}, \ and\ \bibinfo
  {author} {\bibfnamefont {M.~P.}\ \bibnamefont {Valderrama}},\ }\href
  {\doibase 10.1103/PhysRevD.98.054001} {\bibfield  {journal} {\bibinfo
  {journal} {Phys. Rev.}\ }\textbf {\bibinfo {volume} {D98}},\ \bibinfo {pages}
  {054001} (\bibinfo {year} {2018})},\ \Eprint
  {http://arxiv.org/abs/1707.03802} {arXiv:1707.03802 [hep-ph]} \BibitemShut
  {NoStop}%
\bibitem [{\citenamefont {Martinez~Torres}\ \emph {et~al.}(2019)\citenamefont
  {Martinez~Torres}, \citenamefont {Khemchandani},\ and\ \citenamefont
  {Geng}}]{MartinezTorres:2018zbl}%
  \BibitemOpen
  \bibfield  {author} {\bibinfo {author} {\bibfnamefont {A.}~\bibnamefont
  {Martinez~Torres}}, \bibinfo {author} {\bibfnamefont {K.}~\bibnamefont
  {Khemchandani}}, \ and\ \bibinfo {author} {\bibfnamefont {L.-S.}\
  \bibnamefont {Geng}},\ }\href {\doibase 10.1103/PhysRevD.99.076017}
  {\bibfield  {journal} {\bibinfo  {journal} {Phys. Rev.}\ }\textbf {\bibinfo
  {volume} {D99}},\ \bibinfo {pages} {076017} (\bibinfo {year} {2019})},\
  \Eprint {http://arxiv.org/abs/1809.01059} {arXiv:1809.01059 [hep-ph]}
  \BibitemShut {NoStop}%
\bibitem [{\citenamefont {Machleidt}\ \emph {et~al.}(1987)\citenamefont
  {Machleidt}, \citenamefont {Holinde},\ and\ \citenamefont
  {Elster}}]{Machleidt:1987hj}%
  \BibitemOpen
  \bibfield  {author} {\bibinfo {author} {\bibfnamefont {R.}~\bibnamefont
  {Machleidt}}, \bibinfo {author} {\bibfnamefont {K.}~\bibnamefont {Holinde}},
  \ and\ \bibinfo {author} {\bibfnamefont {C.}~\bibnamefont {Elster}},\ }\href
  {\doibase 10.1016/S0370-1573(87)80002-9} {\bibfield  {journal} {\bibinfo
  {journal} {Phys. Rept.}\ }\textbf {\bibinfo {volume} {149}},\ \bibinfo
  {pages} {1} (\bibinfo {year} {1987})}\BibitemShut {NoStop}%
\bibitem [{\citenamefont {Machleidt}(1989)}]{Machleidt:1989tm}%
  \BibitemOpen
  \bibfield  {author} {\bibinfo {author} {\bibfnamefont {R.}~\bibnamefont
  {Machleidt}},\ }\href@noop {} {\bibfield  {journal} {\bibinfo  {journal}
  {Adv. Nucl. Phys.}\ }\textbf {\bibinfo {volume} {19}},\ \bibinfo {pages}
  {189} (\bibinfo {year} {1989})}\BibitemShut {NoStop}%
\bibitem [{\citenamefont {Voloshin}\ and\ \citenamefont
  {Okun}(1976)}]{Voloshin:1976ap}%
  \BibitemOpen
  \bibfield  {author} {\bibinfo {author} {\bibfnamefont {M.}~\bibnamefont
  {Voloshin}}\ and\ \bibinfo {author} {\bibfnamefont {L.}~\bibnamefont
  {Okun}},\ }\href@noop {} {\bibfield  {journal} {\bibinfo  {journal} {JETP
  Lett.}\ }\textbf {\bibinfo {volume} {23}},\ \bibinfo {pages} {333} (\bibinfo
  {year} {1976})}\BibitemShut {NoStop}%
\bibitem [{\citenamefont {Liu}\ \emph {et~al.}(2019)\citenamefont {Liu},
  \citenamefont {Wu}, \citenamefont {Pavon~Valderrama}, \citenamefont {Xie},\
  and\ \citenamefont {Geng}}]{Liu:2019stu}%
  \BibitemOpen
  \bibfield  {author} {\bibinfo {author} {\bibfnamefont {M.-Z.}\ \bibnamefont
  {Liu}}, \bibinfo {author} {\bibfnamefont {T.-W.}\ \bibnamefont {Wu}},
  \bibinfo {author} {\bibfnamefont {M.}~\bibnamefont {Pavon~Valderrama}},
  \bibinfo {author} {\bibfnamefont {J.-J.}\ \bibnamefont {Xie}}, \ and\
  \bibinfo {author} {\bibfnamefont {L.-S.}\ \bibnamefont {Geng}},\ }\href
  {\doibase 10.1103/PhysRevD.99.094018} {\bibfield  {journal} {\bibinfo
  {journal} {Phys. Rev.}\ }\textbf {\bibinfo {volume} {D99}},\ \bibinfo {pages}
  {094018} (\bibinfo {year} {2019})},\ \Eprint
  {http://arxiv.org/abs/1902.03044} {arXiv:1902.03044 [hep-ph]} \BibitemShut
  {NoStop}%
\bibitem [{\citenamefont {Kamimura}(1988)}]{Kamimura:1988zz}%
  \BibitemOpen
  \bibfield  {author} {\bibinfo {author} {\bibfnamefont {M.}~\bibnamefont
  {Kamimura}},\ }\href {\doibase 10.1103/PhysRevA.38.621} {\bibfield  {journal}
  {\bibinfo  {journal} {Phys. Rev.}\ }\textbf {\bibinfo {volume} {A38}},\
  \bibinfo {pages} {621} (\bibinfo {year} {1988})}\BibitemShut {NoStop}%
\bibitem [{\citenamefont {Hiyama}\ \emph {et~al.}(2003)\citenamefont {Hiyama},
  \citenamefont {Kino},\ and\ \citenamefont {Kamimura}}]{Hiyama:2003cu}%
  \BibitemOpen
  \bibfield  {author} {\bibinfo {author} {\bibfnamefont {E.}~\bibnamefont
  {Hiyama}}, \bibinfo {author} {\bibfnamefont {Y.}~\bibnamefont {Kino}}, \ and\
  \bibinfo {author} {\bibfnamefont {M.}~\bibnamefont {Kamimura}},\ }\href
  {\doibase 10.1016/S0146-6410(03)90015-9} {\bibfield  {journal} {\bibinfo
  {journal} {Prog. Part. Nucl. Phys.}\ }\textbf {\bibinfo {volume} {51}},\
  \bibinfo {pages} {223} (\bibinfo {year} {2003})}\BibitemShut {NoStop}%
\bibitem [{\citenamefont {Martinez~Torres}\ \emph
  {et~al.}(2008{\natexlab{a}})\citenamefont {Martinez~Torres}, \citenamefont
  {Khemchandani},\ and\ \citenamefont {Oset}}]{MartinezTorres:2007sr}%
  \BibitemOpen
  \bibfield  {author} {\bibinfo {author} {\bibfnamefont {A.}~\bibnamefont
  {Martinez~Torres}}, \bibinfo {author} {\bibfnamefont {K.~P.}\ \bibnamefont
  {Khemchandani}}, \ and\ \bibinfo {author} {\bibfnamefont {E.}~\bibnamefont
  {Oset}},\ }\href {\doibase 10.1103/PhysRevC.77.042203} {\bibfield  {journal}
  {\bibinfo  {journal} {Phys. Rev.}\ }\textbf {\bibinfo {volume} {C77}},\
  \bibinfo {pages} {042203} (\bibinfo {year} {2008}{\natexlab{a}})},\ \Eprint
  {http://arxiv.org/abs/0706.2330} {arXiv:0706.2330 [nucl-th]} \BibitemShut
  {NoStop}%
\bibitem [{\citenamefont {Khemchandani}\ \emph {et~al.}(2008)\citenamefont
  {Khemchandani}, \citenamefont {Martinez~Torres},\ and\ \citenamefont
  {Oset}}]{Khemchandani:2008rk}%
  \BibitemOpen
  \bibfield  {author} {\bibinfo {author} {\bibfnamefont {K.~P.}\ \bibnamefont
  {Khemchandani}}, \bibinfo {author} {\bibfnamefont {A.}~\bibnamefont
  {Martinez~Torres}}, \ and\ \bibinfo {author} {\bibfnamefont {E.}~\bibnamefont
  {Oset}},\ }\href {\doibase 10.1140/epja/i2008-10625-3} {\bibfield  {journal}
  {\bibinfo  {journal} {Eur. Phys. J.}\ }\textbf {\bibinfo {volume} {A37}},\
  \bibinfo {pages} {233} (\bibinfo {year} {2008})},\ \Eprint
  {http://arxiv.org/abs/0804.4670} {arXiv:0804.4670 [nucl-th]} \BibitemShut
  {NoStop}%
\bibitem [{\citenamefont {Martinez~Torres}\ \emph
  {et~al.}(2008{\natexlab{b}})\citenamefont {Martinez~Torres}, \citenamefont
  {Khemchandani}, \citenamefont {Geng}, \citenamefont {Napsuciale},\ and\
  \citenamefont {Oset}}]{MartinezTorres:2008gy}%
  \BibitemOpen
  \bibfield  {author} {\bibinfo {author} {\bibfnamefont {A.}~\bibnamefont
  {Martinez~Torres}}, \bibinfo {author} {\bibfnamefont {K.~P.}\ \bibnamefont
  {Khemchandani}}, \bibinfo {author} {\bibfnamefont {L.~S.}\ \bibnamefont
  {Geng}}, \bibinfo {author} {\bibfnamefont {M.}~\bibnamefont {Napsuciale}}, \
  and\ \bibinfo {author} {\bibfnamefont {E.}~\bibnamefont {Oset}},\ }\href
  {\doibase 10.1103/PhysRevD.78.074031} {\bibfield  {journal} {\bibinfo
  {journal} {Phys. Rev.}\ }\textbf {\bibinfo {volume} {D78}},\ \bibinfo {pages}
  {074031} (\bibinfo {year} {2008}{\natexlab{b}})},\ \Eprint
  {http://arxiv.org/abs/0801.3635} {arXiv:0801.3635 [nucl-th]} \BibitemShut
  {NoStop}%
\bibitem [{\citenamefont {Martinez~Torres}\ \emph
  {et~al.}(2009{\natexlab{a}})\citenamefont {Martinez~Torres}, \citenamefont
  {Khemchandani},\ and\ \citenamefont {Oset}}]{MartinezTorres:2008kh}%
  \BibitemOpen
  \bibfield  {author} {\bibinfo {author} {\bibfnamefont {A.}~\bibnamefont
  {Martinez~Torres}}, \bibinfo {author} {\bibfnamefont {K.~P.}\ \bibnamefont
  {Khemchandani}}, \ and\ \bibinfo {author} {\bibfnamefont {E.}~\bibnamefont
  {Oset}},\ }\href {\doibase 10.1103/PhysRevC.79.065207} {\bibfield  {journal}
  {\bibinfo  {journal} {Phys. Rev.}\ }\textbf {\bibinfo {volume} {C79}},\
  \bibinfo {pages} {065207} (\bibinfo {year} {2009}{\natexlab{a}})},\ \Eprint
  {http://arxiv.org/abs/0812.2235} {arXiv:0812.2235 [nucl-th]} \BibitemShut
  {NoStop}%
\bibitem [{\citenamefont {Martinez~Torres}\ \emph
  {et~al.}(2009{\natexlab{b}})\citenamefont {Martinez~Torres}, \citenamefont
  {Khemchandani}, \citenamefont {Gamermann},\ and\ \citenamefont
  {Oset}}]{MartinezTorres:2009xb}%
  \BibitemOpen
  \bibfield  {author} {\bibinfo {author} {\bibfnamefont {A.}~\bibnamefont
  {Martinez~Torres}}, \bibinfo {author} {\bibfnamefont {K.}~\bibnamefont
  {Khemchandani}}, \bibinfo {author} {\bibfnamefont {D.}~\bibnamefont
  {Gamermann}}, \ and\ \bibinfo {author} {\bibfnamefont {E.}~\bibnamefont
  {Oset}},\ }\href {\doibase 10.1103/PhysRevD.80.094012} {\bibfield  {journal}
  {\bibinfo  {journal} {Phys.Rev.}\ }\textbf {\bibinfo {volume} {D80}},\
  \bibinfo {pages} {094012} (\bibinfo {year} {2009}{\natexlab{b}})},\ \Eprint
  {http://arxiv.org/abs/0906.5333} {arXiv:0906.5333 [nucl-th]} \BibitemShut
  {NoStop}%
\bibitem [{\citenamefont {Martinez~Torres}\ \emph
  {et~al.}(2009{\natexlab{c}})\citenamefont {Martinez~Torres}, \citenamefont
  {Khemchandani}, \citenamefont {Meissner},\ and\ \citenamefont
  {Oset}}]{MartinezTorres:2009cw}%
  \BibitemOpen
  \bibfield  {author} {\bibinfo {author} {\bibfnamefont {A.}~\bibnamefont
  {Martinez~Torres}}, \bibinfo {author} {\bibfnamefont {K.~P.}\ \bibnamefont
  {Khemchandani}}, \bibinfo {author} {\bibfnamefont {U.-G.}\ \bibnamefont
  {Meissner}}, \ and\ \bibinfo {author} {\bibfnamefont {E.}~\bibnamefont
  {Oset}},\ }\href {\doibase 10.1140/epja/i2009-10834-2} {\bibfield  {journal}
  {\bibinfo  {journal} {Eur. Phys. J.}\ }\textbf {\bibinfo {volume} {A41}},\
  \bibinfo {pages} {361} (\bibinfo {year} {2009}{\natexlab{c}})},\ \Eprint
  {http://arxiv.org/abs/0902.3633} {arXiv:0902.3633 [nucl-th]} \BibitemShut
  {NoStop}%
\bibitem [{\citenamefont {Martinez~Torres}\ and\ \citenamefont
  {Jido}(2010)}]{MartinezTorres:2010zv}%
  \BibitemOpen
  \bibfield  {author} {\bibinfo {author} {\bibfnamefont {A.}~\bibnamefont
  {Martinez~Torres}}\ and\ \bibinfo {author} {\bibfnamefont {D.}~\bibnamefont
  {Jido}},\ }\href {\doibase 10.1103/PhysRevC.82.038202} {\bibfield  {journal}
  {\bibinfo  {journal} {Phys. Rev.}\ }\textbf {\bibinfo {volume} {C82}},\
  \bibinfo {pages} {038202} (\bibinfo {year} {2010})},\ \Eprint
  {http://arxiv.org/abs/1008.0457} {arXiv:1008.0457 [nucl-th]} \BibitemShut
  {NoStop}%
\bibitem [{\citenamefont {Martinez~Torres}\ \emph
  {et~al.}(2011{\natexlab{a}})\citenamefont {Martinez~Torres}, \citenamefont
  {Khemchandani}, \citenamefont {Jido},\ and\ \citenamefont
  {Hosaka}}]{MartinezTorres:2011vh}%
  \BibitemOpen
  \bibfield  {author} {\bibinfo {author} {\bibfnamefont {A.}~\bibnamefont
  {Martinez~Torres}}, \bibinfo {author} {\bibfnamefont {K.~P.}\ \bibnamefont
  {Khemchandani}}, \bibinfo {author} {\bibfnamefont {D.}~\bibnamefont {Jido}},
  \ and\ \bibinfo {author} {\bibfnamefont {A.}~\bibnamefont {Hosaka}},\ }\href
  {\doibase 10.1103/PhysRevD.84.074027} {\bibfield  {journal} {\bibinfo
  {journal} {Phys. Rev.}\ }\textbf {\bibinfo {volume} {D84}},\ \bibinfo {pages}
  {074027} (\bibinfo {year} {2011}{\natexlab{a}})},\ \Eprint
  {http://arxiv.org/abs/1106.6101} {arXiv:1106.6101 [nucl-th]} \BibitemShut
  {NoStop}%
\bibitem [{\citenamefont {Martinez~Torres}\ \emph
  {et~al.}(2011{\natexlab{b}})\citenamefont {Martinez~Torres}, \citenamefont
  {Jido},\ and\ \citenamefont {Kanada-En'yo}}]{Torres:2011jt}%
  \BibitemOpen
  \bibfield  {author} {\bibinfo {author} {\bibfnamefont {A.}~\bibnamefont
  {Martinez~Torres}}, \bibinfo {author} {\bibfnamefont {D.}~\bibnamefont
  {Jido}}, \ and\ \bibinfo {author} {\bibfnamefont {Y.}~\bibnamefont
  {Kanada-En'yo}},\ }\href {\doibase 10.1103/PhysRevC.83.065205} {\bibfield
  {journal} {\bibinfo  {journal} {Phys. Rev.}\ }\textbf {\bibinfo {volume}
  {C83}},\ \bibinfo {pages} {065205} (\bibinfo {year} {2011}{\natexlab{b}})},\
  \Eprint {http://arxiv.org/abs/1102.1505} {arXiv:1102.1505 [nucl-th]}
  \BibitemShut {NoStop}%
\bibitem [{\citenamefont {Faddeev}(1961)}]{Faddeev:1960su}%
  \BibitemOpen
  \bibfield  {author} {\bibinfo {author} {\bibfnamefont {L.~D.}\ \bibnamefont
  {Faddeev}},\ }\href@noop {} {\bibfield  {journal} {\bibinfo  {journal} {Sov.
  Phys. JETP}\ }\textbf {\bibinfo {volume} {12}},\ \bibinfo {pages} {1014}
  (\bibinfo {year} {1961})},\ \bibinfo {note} {[Zh. Eksp. Teor.
  Fiz.39,1459(1960)]}\BibitemShut {NoStop}%
\bibitem [{\citenamefont {Guo}\ \emph {et~al.}(2007)\citenamefont {Guo},
  \citenamefont {Shen},\ and\ \citenamefont {Chiang}}]{Guo:2006rp}%
  \BibitemOpen
  \bibfield  {author} {\bibinfo {author} {\bibfnamefont {F.-K.}\ \bibnamefont
  {Guo}}, \bibinfo {author} {\bibfnamefont {P.-N.}\ \bibnamefont {Shen}}, \
  and\ \bibinfo {author} {\bibfnamefont {H.-C.}\ \bibnamefont {Chiang}},\
  }\href {\doibase 10.1016/j.physletb.2007.01.050} {\bibfield  {journal}
  {\bibinfo  {journal} {Phys. Lett.}\ }\textbf {\bibinfo {volume} {B647}},\
  \bibinfo {pages} {133} (\bibinfo {year} {2007})},\ \Eprint
  {http://arxiv.org/abs/hep-ph/0610008} {arXiv:hep-ph/0610008 [hep-ph]}
  \BibitemShut {NoStop}%
\bibitem [{\citenamefont {Lang}\ \emph {et~al.}(2015)\citenamefont {Lang},
  \citenamefont {Mohler}, \citenamefont {Prelovsek},\ and\ \citenamefont
  {Woloshyn}}]{Lang:2015hza}%
  \BibitemOpen
  \bibfield  {author} {\bibinfo {author} {\bibfnamefont {C.~B.}\ \bibnamefont
  {Lang}}, \bibinfo {author} {\bibfnamefont {D.}~\bibnamefont {Mohler}},
  \bibinfo {author} {\bibfnamefont {S.}~\bibnamefont {Prelovsek}}, \ and\
  \bibinfo {author} {\bibfnamefont {R.~M.}\ \bibnamefont {Woloshyn}},\ }\href
  {\doibase 10.1016/j.physletb.2015.08.038} {\bibfield  {journal} {\bibinfo
  {journal} {Phys. Lett.}\ }\textbf {\bibinfo {volume} {B750}},\ \bibinfo
  {pages} {17} (\bibinfo {year} {2015})},\ \Eprint
  {http://arxiv.org/abs/1501.01646} {arXiv:1501.01646 [hep-lat]} \BibitemShut
  {NoStop}%
\end{thebibliography}%

\end{document}